# Size scaling, dynamics, and electro-thermal bifurcation of $VO_2$ Mott oscillators

Stephanie M. Bohaichuk,[1*] Suhas Kumar,[2*] Miguel Muñoz Rojo,[1,3] R. Stanley Williams,[4] Mahnaz Islam,[1] Gregory Pitner,[1] Jaewoo Jeong,[5] Mahesh G. Samant,[5] Stuart S. P. Parkin,[5] Eric Pop[1,6]

[1]*Dept. of Electrical Engineering, Stanford University, Stanford, CA 94305, USA*
[2]*Hewlett Packard Labs, 1501 Page Mill Rd, Palo Alto, CA 94304, USA*
[3]*Dept. of Thermal and Fluid Engineering, University of Twente, 7500 AE Enschede, The Netherlands*
[4]*Dept. of Electrical & Computer Engineering, Texas A&M University, College Station, TX 77843, USA*
[5]*IBM Almaden Research Center, 650 Harry Rd, San Jose, CA 95120, USA*
[6]*Dept. of Materials Science and Engineering, Stanford University, Stanford, CA 94305, USA*
**Email: sbohaich@stanford.edu (S.M.B.), su1@alumni.stanford.edu (S.K.)*

**Traditional electronic devices are well-known to improve in speed and energy-efficiency as their dimensions are reduced to the nanoscale. However, this scaling behavior remains unclear for nonlinear dynamical circuit elements, such as Mott neuron-like spiking oscillators, which are of interest for bio-inspired computing. Here we show that shrinking micrometer-sized $VO_2$ oscillators to sub-100 nm effective sizes, achieved using a nanogap cut in a metallic carbon nanotube (CNT) electrode, does not guarantee faster spiking. However, an additional heat source such as Joule heating from the CNT, in combination with small size and heat capacity (defined by the narrow volume of $VO_2$ whose insulator-metal transition is triggered by the CNT), can increase the spiking frequency by ~1000× due to an electro-thermal bifurcation in the nonlinear dynamics. These results demonstrate that nonlinear dynamical switches operate in a complex phase space which can be controlled by careful electro-thermal design, offering new tuning parameters for designing future biomimetic electronics.**

Brain-inspired computing systems have gained considerable interest as a path to more efficient machine learning and artificial intelligence. Proposed hardware implementations of these systems often require fast, controllable, low energy sources of neuron-like spiking [1,2]. In typical transistor-based circuits, this behavior cannot be achieved in a single device, requiring a more elaborate circuit [3]. Instead, spiking can be produced in compact Mott memristive switches, which are two-terminal devices exhibiting electronic instabilities such as negative differential resistance (NDR), often constructed using Mott insulators that undergo an insulator-metal transition (IMT) [4-6]. These include vanadium dioxide ($VO_2$) and niobium dioxide ($NbO_2$), which undergo an IMT at ~340 K and ~1070 K, respectively. Above this IMT temperature, the material undergoes an abrupt increase in conductivity, typically by several orders of magnitude, which reverses upon cooling.

Devices constructed from such materials exhibit strong coupling between nonlinear thermally activated electrical transport (especially in the case of an IMT) and localized Joule heating [7,8]. When measured using a voltage source, feedback between these processes leads to abrupt volatile switching in current and resistance, but when a device is driven by a current source, it displays an electronic instability and NDR.

When biased within the region of NDR in combination with a parallel capacitor(s) (whether externally added, parasitic, or intrinsic to the device), the volatile resistive switching of the device, together with the parallel capacitor's charging and discharging, can produce periodic self-sustained oscillations in the device voltage and current. This setup is known as a relaxation oscillator or a Pearson-Anson oscillator, and can produce sharp neuron-like spiking, useful in the construction of biomimetic circuits [9-13].

An increase in oscillation frequency and a reduction in energy might logically be gained by shrinking Mott switches down to the nanoscale, especially if the device capacitance reduces [5,14], similar in principle to traditional transistor scaling. The incubation (delay) time needed to electrically trigger switching from the insulating to the metallic state is known to scale with the length and width of Mott switches [15]. However, an understanding of scaling and methods for engineering the full dynamical spiking behavior in these non-linear devices have not been sufficiently developed, and such an understanding is essential for designing compact biomimetic hardware [16].

In this work, we probed the IMT in sub-100 nm regions of $VO_2$ using a nanogap cut in a single-wall metallic carbon nanotube (CNT), which formed ultra-narrow electrodes with ~1 nm diameter. This nanogap test platform was inspired by previous work with phase-change memory [17], but has not been applied to volatile switches until now. Despite the extremely small volume (low thermal mass) heated and cooled across the IMT, these nanogap devices oscillated at comparable frequencies (~kHz) to much larger, micrometer-sized devices. However, when a continuous single metallic carbon nanotube [18] was used as an additional nanoscale heat source (or alternatively, a lithographically defined heater line or conductive bilayer structure, as demonstrated elsewhere [19,20]) the frequency of oscillations dramatically increased (to ~MHz), accompanied by a reduction in the energy per spike delivered to a load. We show that these nonlinear devices are susceptible to abrupt shifts in dynamical behavior, known as bifurcations, as a function of changes in electro-thermal parameters, which can be exploited as additional design parameters.

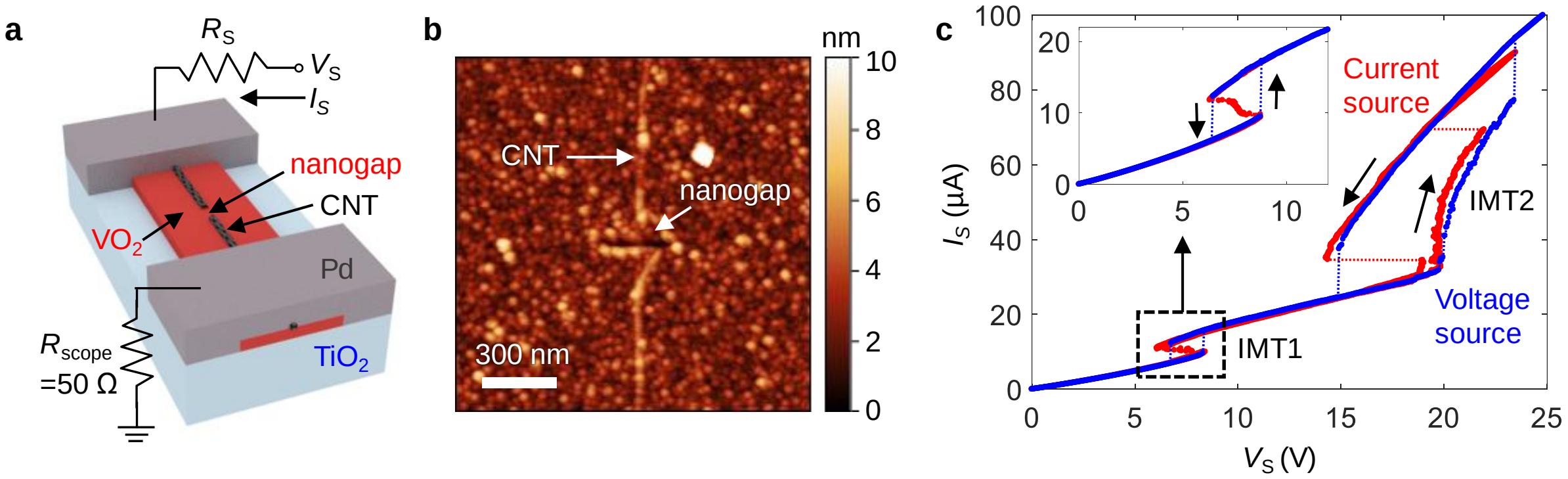

**FIG. 1.** CNT-$VO_2$ nanogap structure and static behavior. (a) Schematic illustration and measurement setup for a $VO_2$ device with a cut metallic CNT on top forming the electrodes. $R_{scope}$ is the resistance of the measurement oscilloscope and $R_S$ = 200 kΩ is a series resistor. (b) AFM image of a CNT on $VO_2$ after cutting the CNT with an AFM tip, creating a <100 nm nanogap. (c) Typical quasi-static current-voltage characteristics of a nanogap device measured using a voltage source (blue) and a current source (red), showing that switching occurred in two steps. The first corresponded to the IMT of the nano-sized $VO_2$ volume within the CNT gap (IMT1), and the second to the larger IMT volume of $VO_2$ between the Pd metal contacts along the CNT (IMT2), which was connected by metallic (post-IMT1) $VO_2$ within the nanogap. The inset magnifies IMT1 with identical units on the axes.

As shown in Figures 1a-b, we fabricated nanoscale $VO_2$ devices by utilizing single-walled metallic CNTs with ~1 nm diameter as electrodes. Aligned CNTs were grown on a quartz substrate and transferred [21,22] onto a thin film (thickness 5 nm) of $VO_2$ grown epitaxially [23] on $TiO_2$ (101). After patterning the $VO_2$ by wet etching, 50 nm thick Pd was deposited to make metallic contacts to both the $VO_2$ and CNTs [22]. A CNT running between the contacts was physically cut near its midpoint by using an atomic force microscope (AFM) tip. (See Section 1 of the Supplemental Material [24] for additional fabrication details including lithography, etching, and AFM cutting.)

The current-voltage behavior of the nanogap device (Figure 1c, corresponding to the device in Figure 1b) at lower voltages (<10 V) was dominated by the insulating $VO_2$ in the nanogap, with an IMT occurring in the gap at $V_S$ = 8.3 V (abrupt jump in current marked 'IMT1'). Once the $VO_2$ in the gap became metallic, as $V_S$ was increased, the now-metallically-connected CNT acted as a localized Joule heater (i.e., the gap in the CNT was shorted by the metallic $VO_2$ within). This caused a second, larger IMT (marked 'IMT2') in the $VO_2$ along the length of the CNT, associated with a region of NDR (seen once the voltage drop across $R_S$ is subtracted, see Supplemental Material Figure S2 [24]) between $V_S$ = 20 V and 23.4 V, followed by an abrupt jump in current at ~23.4 V. The IMT2 behavior is similar to that observed in previous work with a continuous (uncut) CNT as the heater [22]. Switching was repeatable and similar among other CNT-$VO_2$ nanogap devices, although the switching voltage and magnitude of change in resistance were dependent on the nanogap length (see Sections 2 and 4 of the Supplemental Material [24]).

To experimentally validate the heated volume and switching region associated with each IMT in these devices, we used Kelvin Probe Microscopy (KPM) and Scanning Thermal Microscopy (SThM), two scanning probe techniques with sub-100 nm spatial resolution. KPM maps the local surface potential in a biased device, and when the image is flattened (i.e., by removing the average linear potential drop between metal contacts, see Section 3 of the Supplemental Material [24]), regions of contrast highlight changes in resistance and electric field [22,25]. KPM of a nanogap device held at $V_S$ = 8 V, just prior to IMT1, exhibited sharp contrast at the CNT nanogap (Figure 2a), indicating a concentrated field within it. SThM was used to map changes in device heating [22,26], as seen on the surface of a 35 nm poly(methyl methacrylate) (PMMA) capping layer used to electrically isolate the SThM tip from the device. The same device at an identical bias imaged using SThM exhibited localized heating within the nanogap (Figure 2b), while no significant heating was observed along the rest of the CNT or $VO_2$. This was consistent with finite-element electrothermal simulations (Figure 2c) based on a thermally-induced IMT in the nanogap (see Section 4 of the Supplemental Material [24]), which showed that significant heating occurred only in the $VO_2$ within the gap and at the interface between the $VO_2$ and cut CNT ends. Heating at the two CNT tips forming the nanogap was dominant in the simulation, but blurring due to the finite tip size and thermal exchange radius [27] can cause the experimental appearance of a single hot spot centered on the gap. Just after IMT1, the simulated maximum temperature of the $VO_2$ in the nanogap was only ~10 K above its IMT temperature ($T_{IMT}$ = 328 K) in steady state.

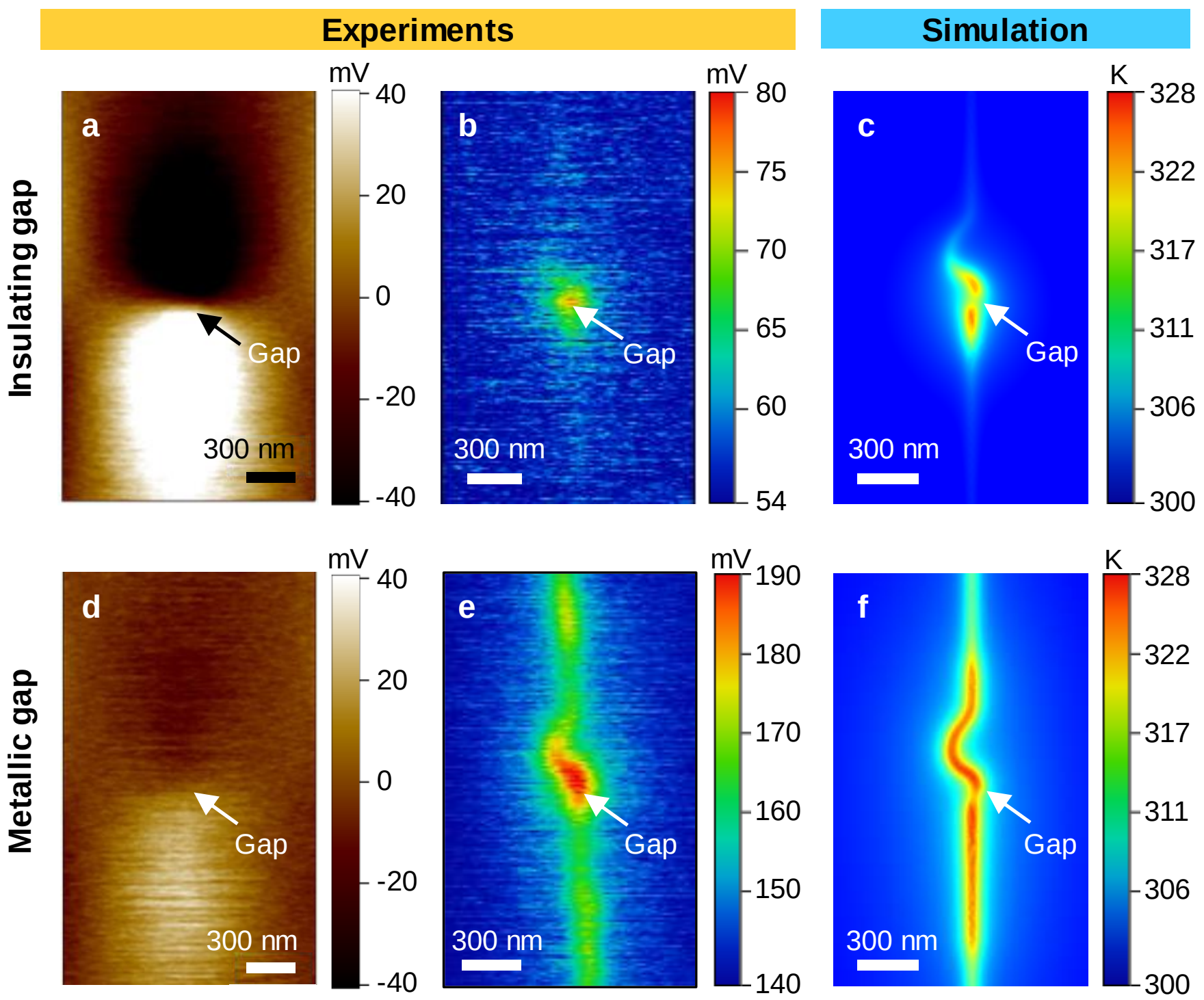

**FIG. 2.** Localized electric field and temperature during IMT1 and IMT2. (a) Flattened KPM image of a nanogap device held at $V_S = 8$ V (with $R_S = 200$ kΩ), just prior to IMT1. First-order flattening of the surface potential removes the average linear potential drop between the Pd contacts, and the contrast at the gap indicates a strong field within it [24]. The positive contact (ground) is outside the bottom (top) of each image. (b) SThM image of the same bias in the same device (now capped by PMMA) showing highly localized heating in the gap. The color bar refers to the SThM voltage, which is a measure of the change in local temperature on the PMMA surface. (c) Simulation of the capped nanogap device, showing the temperature on the PMMA surface prior to IMT1. (d) Flattened KPM image of the same device at $V_S = 16$ V, after IMT1 but before IMT2. There is a lower potential drop across the gap once it is metallic. (e) SThM image of the same bias and device (capped by PMMA), showing heating in the gap and also along the rest of the now-connected CNT. (f) Corresponding simulation of the capped device temperature on the PMMA surface, after IMT1.

Beyond IMT1 but prior to inducing IMT2 (at $V_S = 16$ V), much weaker contrast was observed in the flattened KPM image (Figure 2d), consistent with a lower voltage drop and field across the $VO_2$ in the gap, which had turned metallic after IMT1. SThM (Figure 2e) at this bias indicated that heating occurred not only in the gap, but also along the full length of the CNT, indicating that the CNT was effectively reconnected by the metallic $VO_2$ in the gap. Finite element simulations (Figure 2f) also confirmed heating of the $VO_2$ along the entire CNT length, leading to IMT2.

When biased with a constant current within a region of NDR, it is possible to produce self-sustained periodic electrical oscillations, aided by a parallel capacitor that is often intrinsic to the device [10,18, 28]. In a comparable micrometer-scale $VO_2$ device made *without* a CNT ($L = 3.5$ µm, $W = 2.7$ µm) oscillations occurred with a frequency of ~0.4 kHz (Figure 3a), consisting of a fast (~70 ns) initial spike (inset of Figure 3d) followed by a slow decay of ~0.44 ms. In the nanogap device of Figure 1, oscillations occurred at IMT1 (corresponding to the nanogap volume) with a frequency of ~0.6 kHz (Figure 3b). Thus, despite an enormous reduction in the volume of $VO_2$ heated and cooled across the IMT (observed in Figures 2a-2c), the CNT nanogap device oscillated at nearly the same frequency as the large $VO_2$-only device. Similar slow oscillations were observed in other nanogap devices (see Section 2b of the Supplemental Material [24]). This suggests that merely reducing the switching volume of $VO_2$ may not be sufficient for high-frequency spiking. The spiking frequency of IMT devices may often be limited by parasitic capacitances (externally or as part of the device contacts or geometry), or in some cases by the current source itself (see Section 2c of the Supplemental Material [24] for possible influence of the current source on oscillation frequency), rather than solely from device size and thermal mass.

However, when the same CNT nanogap device was biased within IMT2, the CNT was re-connected as a Joule heating source (in series with the metallic $VO_2$ bridging the gap) and oscillations were observed to be over 1000 times faster with a frequency of 0.65 MHz (Figures 3c-3f). This is surprising because the thermal volume for the IMT2 'connected nanogap' was much larger than for the IMT1 nanogap (evidenced by Figure 2), and a larger thermal volume usually implies slower dynamics. The measurement setup (including current source, cables, probes) and device are the same as in IMT1, meaning similar parasitics are also expected. Yet the connected nanogap device was orders of magnitude faster.

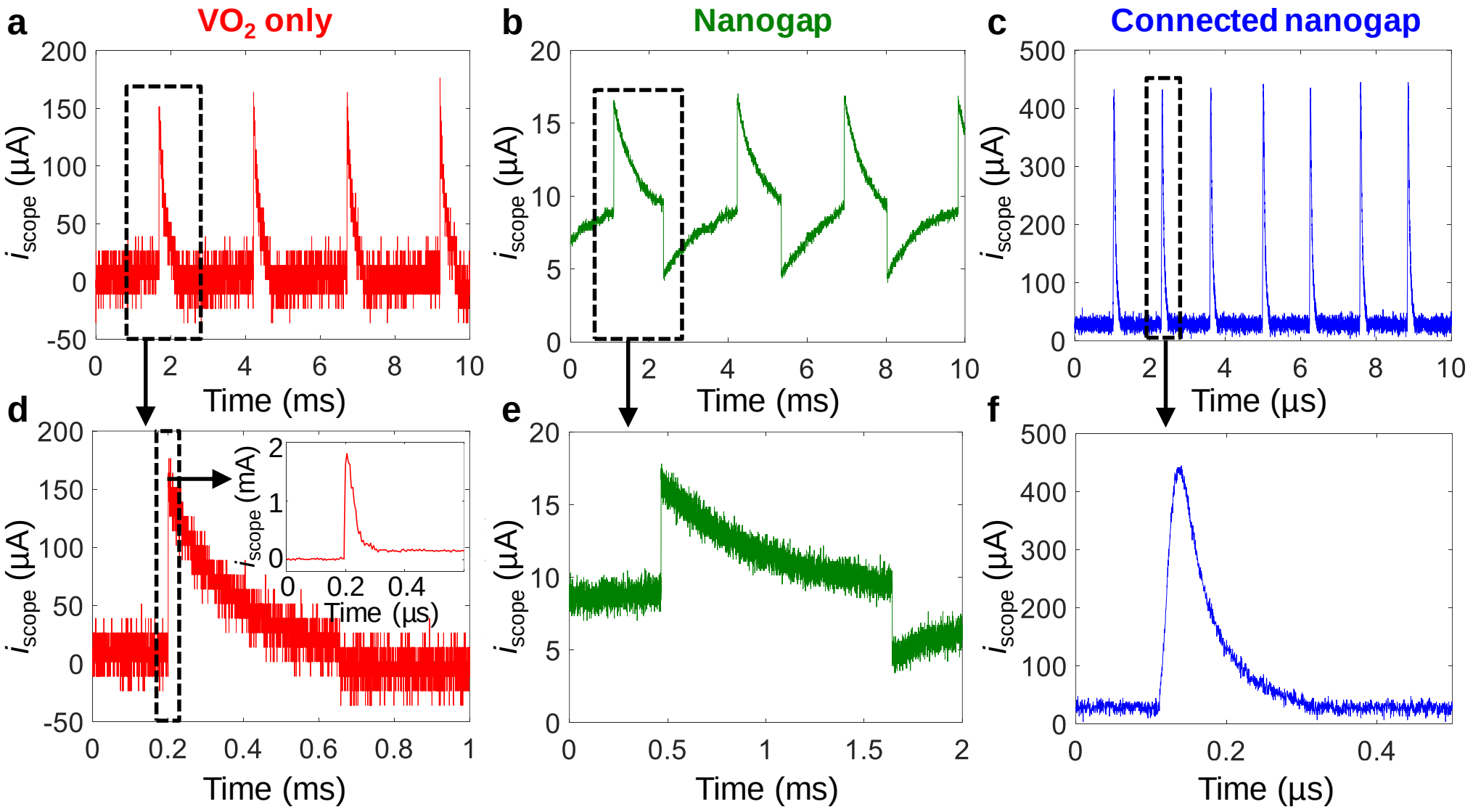

**FIG. 3.** Dynamics in $VO_2$ devices with and without a nanogap. Oscillations were measured with a 50 Ω oscilloscope in series, when the device was biased with a constant current in a region of negative differential resistance (NDR). (a) A micrometer-scale $VO_2$ device ($L$ = 3.5 μm, $W$ = 2.7 μm) without a CNT, corresponding to oscillations of a large volume of $VO_2$. (b) A CNT nanogap device biased in IMT1, corresponding to oscillations of a nanoscale volume of $VO_2$ in the gap ($L_{gap}$ < 100 nm). (c) A CNT nanogap device biased in IMT2, corresponding to oscillations of $VO_2$ in a narrow region below the CNT, but extending along its full length ($L$ = 3.4 μm). Despite the larger volume heated and cooled across the IMT, the $VO_2$ during IMT2 oscillated ~1000× faster than in IMT1. (d)-(f) are magnified plots of (a)-(c).

To gain insight into these results we constructed a compact model, which reveals that nonlinear electronic switching devices are susceptible to abrupt changes in oscillation frequency and dynamical behavior, known as bifurcations, as a function of an appropriate tuning parameter. Here, we observed that the thermal capacitance and the strength of an added heat source can be tuning parameters in such thermally-driven nonlinear devices.

The model consisted of nonlinear thermally-activated Schottky transport for the device, coupled to Newton's law of cooling (Equation 1), which describes the competition between self-heating and heat loss to the environment. $C_{th}$ is the thermal capacitance (which scales with switching volume), $R_{th}$ is a lumped thermal resistance between the hot device and the environment, $T$ represents an average device temperature, and $T_0$ is the ambient temperature. $i_m$ and $v_m$ are the current through and voltage across the device, respectively. We examined the effect of scaling $C_{th}$, as well as the effect of including a resistive heater ($R_{heater}$), such as the CNT which appeared electrically in parallel to the oscillating $VO_2$ biased at IMT2. This heater added a Joule heat source term in the thermal dynamics of Equation 1.

$$C_{\mathrm{th}} \frac{\mathrm{d}T}{\mathrm{d}t} = i_{\mathrm{m}} v_{\mathrm{m}} + \frac{v_{\mathrm{m}}^2}{R_{heater}} - \frac{T - T_0}{R_{\mathrm{th}}} \tag{1}$$

This nonlinear dynamical model produced abrupt resistive switching during a DC voltage sweep (due to positive feedback and thermal runaway) and an NDR instability during a current sweep, similar to IMT switching devices. The model approximated the device behavior with mathematically simple Schottky transport, as any sufficiently nonlinear thermal transport will produce volatile switching and NDR (and the associated self-oscillations in a relaxation circuit), although it did not capture all underlying physics. We did not include an explicit IMT mechanism, instead capturing the switching and NDR with a simpler model in order to clearly isolate the effects of tuning parameters that led to dramatic changes in the dynamics and the introduction of bifurcations. The model is detailed in Section 5 of the Supplemental Material [24].

The switching device model was incorporated into a relaxation oscillator circuit including several capacitors (representing intrinsic and parasitic capacitances) and a series resistor (see Section 5 of the Supplemental Material [24]). When held at a constant current in the NDR region with only self-heating present, the circuit simulation reproduced the shape and ~kHz frequency of the oscillations in a simple $VO_2$ device [18]. To simulate the reduction of switching volume from micrometers to tens of nanometers, we reduced $C_{th}$ by a factor of ~1000, but found that the oscillation frequency remained nearly the same (Figure 4a). However, adding a sufficiently conductive Joule heater was found to induce a bifurcation in which the oscillations abruptly sped up to ~MHz, once $C_{th}$ was sufficiently small (Figure 4b). When a Joule heater was either not present or insufficient to cause a bifurcation, the oscillation frequency remained slow over a large range of $C_{th}$ (Figure 4c). These simulations agree well with the experimental data.

Thus, in Mott switches, the thermal capacitance $C_{th}$ acts as a tuning parameter in the system for a fixed, sufficiently small $R_{heater}$, inducing a sudden change in the dynamical behavior below a critical value (Figure 4c). Similarly, the resistive heater (here the CNT) can act as a tuning parameter to abruptly speed up oscillations at a fixed, sufficiently small $C_{th}$ (see Figure S23 in the Supplemental Material [24]). These results show that volume scaling can be important for achieving faster oscillations by determining $C_{th}$, but must be accompanied by a careful design of the circuit and analysis of the nonlinear dynamics, which can give rise to abrupt bifurcations and orders-of-magnitude changes in performance.

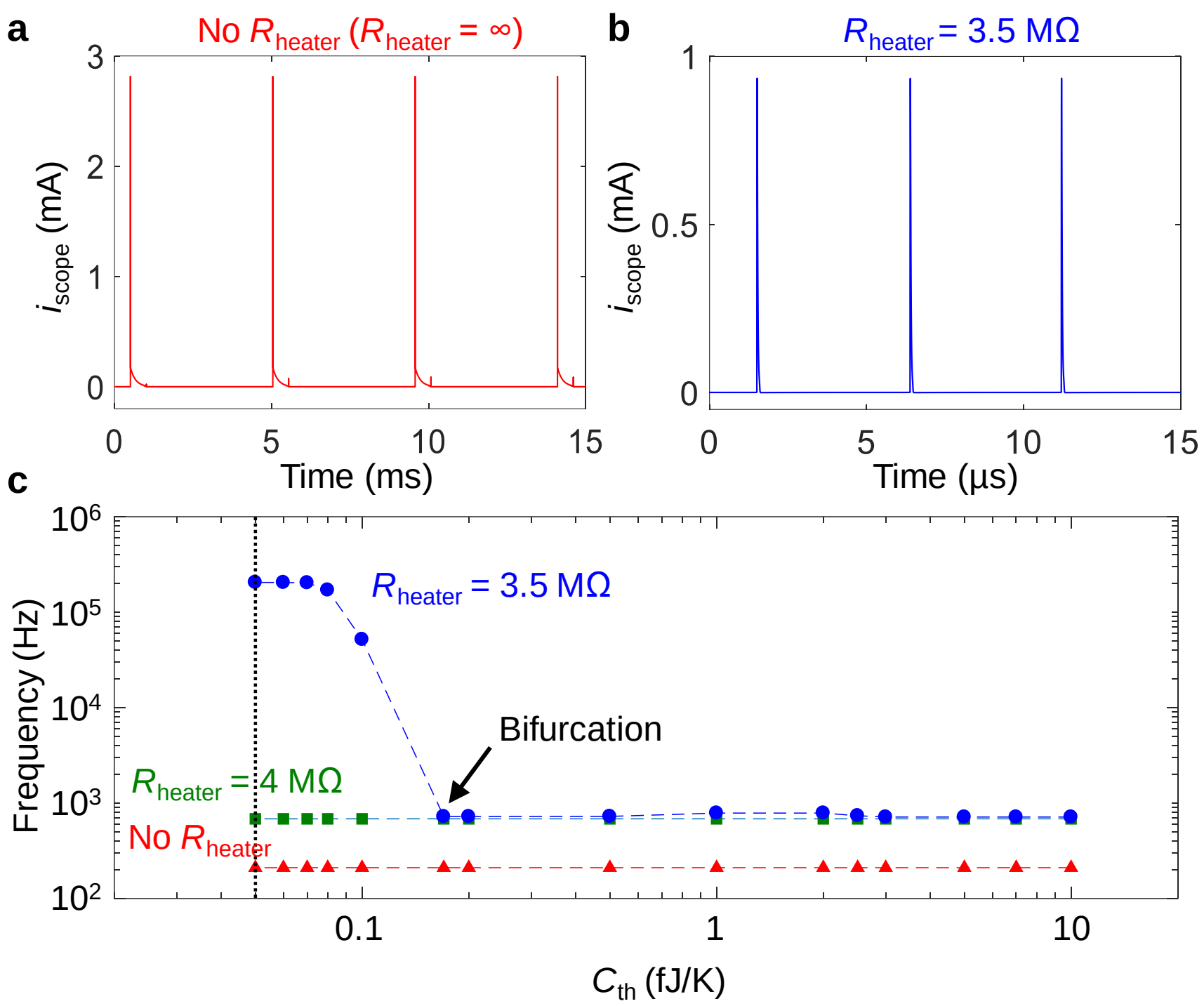

**FIG. 4.** Modeling of bifurcations. Simulated oscillations based on a nonlinear thermally-driven instability. (a) Oscillations of a simple $VO_2$ device with reduced volume represented by a reduced $C_{th}$. The oscillation frequency is similar to a device with 1000x larger $C_{th}$ [18]. (b) Oscillations of a device with a resistive heater in addition to a reduced $C_{th}$, exhibiting much faster oscillations. (c) Bifurcation plot showing that adding a <3.5 MΩ resistive heater and simultaneously reducing $C_{th}$ produced an abrupt frequency increase (blue). If the device only self-heats (red) or a slightly weaker heater (insufficient to cause a bifurcation) was used (green), then $C_{th}$ had no effect on frequency. The plots in (a) and (b) correspond to the red and blue points on the vertical dashed line in (c), with $C_{th}$ = 0.05 fJ/K.

In conclusion, our results show that the spiking speed of a $VO_2$-based Mott oscillator is not always strictly determined by size alone, but can be driven by bifurcations in the nonlinear electro-thermal coupling. Careful control of tuning parameters, particularly within the device's thermal dynamics, alongside purely electronic design, may often be necessary in addition to size scaling to achieve faster oscillations. Thus, bifurcations in the thermal dynamics of nonlinear circuit elements are a new set of design considerations that were not necessarily relevant for circuit design in the era of Moore's law, but one that future beyond-CMOS nanoelectronics could encounter. This work highlights such understanding, which is important for future design of biomimetic electronic chips.

## Acknowledgements

H. S. Philip Wong is gratefully acknowledged for providing technical advice. Device fabrication was performed at the Stanford Nanofabrication Facility and the Stanford Nano Shared Facilities, supported by the National Science Foundation (NSF) under award ECCS-1542152. This work was supported in part by ON Semiconductor, by the Stanford SystemX Alliance, and by the X-Grants Program of the President's Excellence Fund at Texas A&M University. S.B. acknowledges support from the Stanford Graduate Fellowship (SGF) program and the NSERC Postgraduate Scholarship program.

# Supplemental Material

## Size scaling, dynamics, and electro-thermal bifurcation of $VO_2$ Mott oscillators

Stephanie M. Bohaichuk,[1*] Suhas Kumar,[2*] Miguel Muñoz Rojo,[1,3] R. Stanley Williams,[4] Mahnaz Islam,[1] Gregory Pitner,[1] Jaewoo Jeong,[5] Mahesh G. Samant,[5] Stuart S. P. Parkin,[5] Eric Pop[1,6]

[1]*Dept. of Electrical Engineering, Stanford University, Stanford, CA 94305, USA*
[2]*Hewlett Packard Labs, 1501 Page Mill Rd, Palo Alto, CA 94304, USA*
[3]*Dept. of Thermal and Fluid Engineering, University of Twente, 7500 AE Enschede, The Netherlands*
[4]*Dept. of Electrical & Computer Engineering, Texas A&M University, College Station, TX 77843, USA*
[5]*IBM Almaden Research Center, 650 Harry Rd, San Jose, CA 95120, USA*
[6]*Dept. of Materials Science and Engineering, Stanford University, Stanford, CA 94305, USA*
*Email: sbohaich@stanford.edu (S.M.B.), su1@alumni.stanford.edu (S.K.)*

**Supplemental Material Content:**

## 1. Device Fabrication

A finished device schematic is shown in Figure 1a of the main text. To achieve this device structure, thin films of vanadium dioxide ($VO_2$) were first grown epitaxially on single crystal $TiO_2$ (101) substrates by pulsed laser deposition [1]. Films were smooth (<1 Å rms roughness), with a slight self-limiting surface oxidation to $V_2O_5$. The $VO_2$ exhibited transition temperatures of 328 K during heating (insulator to metal, $T_{IMT}$) and 321 K during cooling (metal to insulator, $T_{MIT}$), with a change in resistance over three orders of magnitude [2].

Separately, carbon nanotubes (CNTs) were grown by chemical vapour deposition on an ST-cut quartz substrate using Fe nanoparticles as catalysts, in order to achieve horizontal alignment of the CNTs [3]. The resulting CNTs were a mixture of metallic and semiconducting, with an average diameter of 1.2 nm and an average density of 1 CNT per 3 µm. The CNTs were transferred onto the surface of the $VO_2$, using 100 nm e-beam evaporated Au as a sacrificial support layer [2,3]. The Au/CNTs were peeled off the quartz using thermal release tape, and then pressed onto the $VO_2$. The tape was removed at 130ºC; then Ar and $O_2$ plasma cleans were done to remove any tape residue off the Au. Finally, the Au was wet etched in a KI solution to leave aligned CNTs on the $VO_2$. Some carbon-based residue was left behind in the process. (See also Ref. [2])

The $VO_2$ was patterned into stripes of width $W$ = 2 to 10 µm, with CNTs outside the patterned stripes removed by a light $O_2$ plasma prior to wet etching the $VO_2$ in 25% nitric acid [2]. Contacts with spacing $L$ = 3 to 8 µm were made of 50 nm e-beam evaporated Pd (no adhesion layer was used) patterned using lift-off. The CNT and the $VO_2$ were electrically in parallel, sharing the same contacts. After processing, the $VO_2$ thickness was measured to be ~ 5 nm.

Devices with a metallic CNT were identified electrically (a $VO_2$ device with a metallic CNT will carry noticeably more current than a device with insulating $VO_2$ alone) [2], and the number of CNTs was confirmed using atomic force microscopy (AFM). Devices with a single metallic CNT were the focus of this study, as schematically shown in Figure 1a in the main text.

### a. How to Cut a CNT

To create nanogaps (see Figures 1 and S1), a metallic CNT within a device was cut using the force of a sharp AFM tip in lithography mode on a Park XE-100 or an Asylum Research MFP-3D system. Lithography mode is a variation of contact mode with a higher setpoint (a higher force applied to the tip), used to locally indent or scratch the sample rather than image it. CNTs were mechanically cut to form the nanogap rather than patterned via e-beam lithography and etched (i.e., via Ar or $O_2$ plasma) in order to avoid damage or stoichiometry changes to the underlying $VO_2$.

For increased hardness and durability, either diamond tips (D300 from K-Tek Nanotechnology, ~40 N/m spring constant, ~300 kHz resonant frequency, 5-10 nm radius), or diamond-like-carbon (DLC) coated Si tips (TAP300DLC from Budget Sensors, ~40 N/m spring constant, ~300 kHz resonant frequency, <15 nm radius) were used.

Care was taken to ensure that the setpoint was high enough to cut the CNT but not so high as to deeply scratch the $VO_2$ underneath. For cuts, a setpoint of 0.7 V was used with a speed of 0.2 – 20 µm/s. Because the CNTs were only held to the $VO_2$ surface by Van der Waals forces, they were susceptible to being "dragged" by the tip rather than cut. The two most successful methods of cutting a CNT were:

i) By using several small back-and-forth cuts in rapid sequence (a sawing action) with high force. Cut lines should be very short to minimize lateral force and drag distance. This was used for the devices shown in Figures 1 and S1b.

ii) By first “stabbing” the CNT (performing a force-distance curve), using a high setpoint such as 1.8 – 2 V. This does not usually break the CNT but weakens it. A cut line can then be done at or just next to the indent point, causing the CNT to snap. This was used for the device in Figure S1a and S1c.

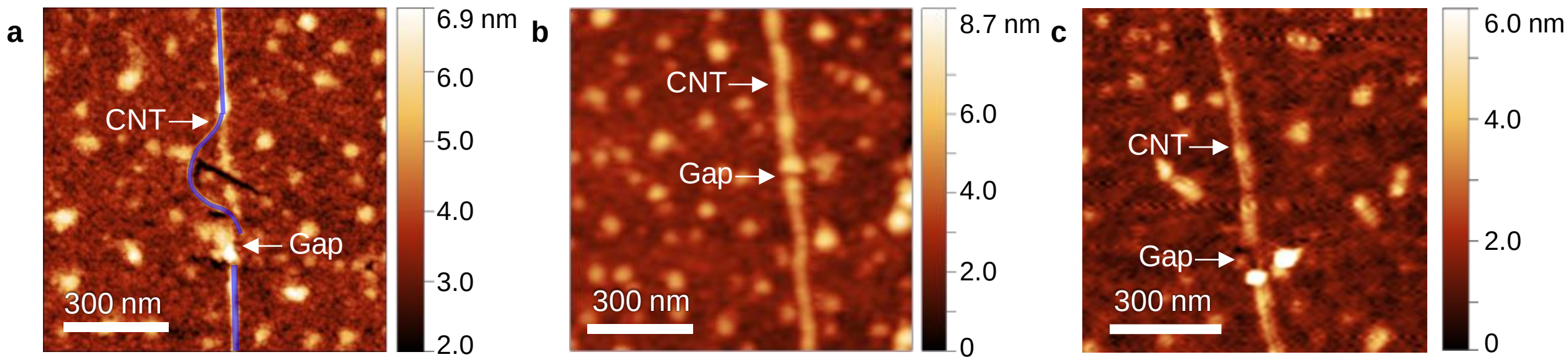

**Figure S1.** (a) A cut CNT (marked in blue) within a $VO_2$ device ($L$ = 4.7 µm, $W$ = 5.0 µm), made by first weakening the CNT and then pulling on it to snap it (there is a scratch in the $VO_2$ near the bent part of the CNT from the AFM tip scan line used to pull on the CNT). (b) A cut CNT within a $VO_2$ device ($L$ = 3.7 µm, $W$ = 7.2 µm) made with a small cutting action of the AFM tip. (c) a cut CNT within a $VO_2$ device ($L$ = 6.5 µm, $W$ = 5.5 µm) made by repeatedly “stabbing” the CNT. The tall particles are carbon-based residue leftover from the CNT transfer process.

### b. How Not to Cut a CNT

Small nanogaps (20 – 300 nm) can be created in metallic carbon nanotubes by electrical breakdown [4], on insulating $SiO_2$ (on Si) substrates. When sufficient bias is applied to the CNT, self-heating causes the CNT to reach its oxidation temperature ~ 600ºC [5]. Since the temperature is not uniform along the CNT (it is hottest in the middle of its length or at a defect site) a gap can be created at a hot spot, and is highly localized if done in an environment with reduced but non-zero oxygen content.

However, the IMT temperature for $VO_2$ (~55ºC) is far lower than 600ºC, and therefore devices switch before the CNT gets hot enough to oxidize. Because of the series resistor (current compliance) used, which is necessary to prevent total device failure, the voltage across the CNT abruptly reduces once IMT occurs. The CNT is thus never able to reach a high enough voltage and therefore temperature to form a nanogap by oxidation.

We also tried forming gaps in CNTs by doing local oxidation lithography. This is a scanning probe technique in which a biased AFM tip induces oxidation underneath the tip by a reaction of the sample with adsorbed water normally present on sample surfaces in air [6]. Oxidation was done at a humidity of 75 – 80%, a tip bias of around -1.9 V, and a setpoint of 0.3 V. However, it was found that $VO_2$ was more susceptible to oxidation than CNTs, and a line of oxide would be formed on the $VO_2$ before a clear gap was made in the CNT. In other words, at best the CNT simply became more resistive rather than broken.

## 2. Electrical Measurements

Electrical measurements were performed using a Keithley 4200-SCS parameter analyzer in a Janis Research probe station (ST-100-UHT-4). $VO_2$ devices with no CNT or with a continuous uncut CNT can be measured equally well in air or vacuum. However, $VO_2$ devices with a cut CNT had to be measured in vacuum (<$10^{-4}$ Torr), as the defective cut ends are more susceptible to oxidation. Unless otherwise specified all measurements were performed at an ambient temperature $T_0$ = 295 K.

For all measurements, including scanning probe measurements, a 200 kΩ resistor was used in series with the device as a current compliance in the metallic state (to avoid overheating and failure). This resistor was added on the probe positioner, right next to the probe tip [2], as close as possible to the device to minimize capacitive current overshoot [7].

## a. DC Electrical Characterization

Measurements of the device in main text Figure 1 before and after cutting the metallic CNT are shown in Figure S2. Before cutting, the CNT acts as a nanoscale heater triggering a single IMT event in the $VO_2$ directly underneath. After the nanogap is formed in the CNT, another earlier IMT (IMT1) is introduced, corresponding to IMT of the $VO_2$ within the nanogap. The second IMT (IMT2) occurs as for the uncut CNT, and corresponds to the IMT of the entire $VO_2$ "sleeve" below the CNT. However, the total device current is reduced, and the switching voltage increased slightly, due to the added resistance of the $VO_2$ in the nanogap, and the resistance of the CNT-$VO_2$ interfaces.

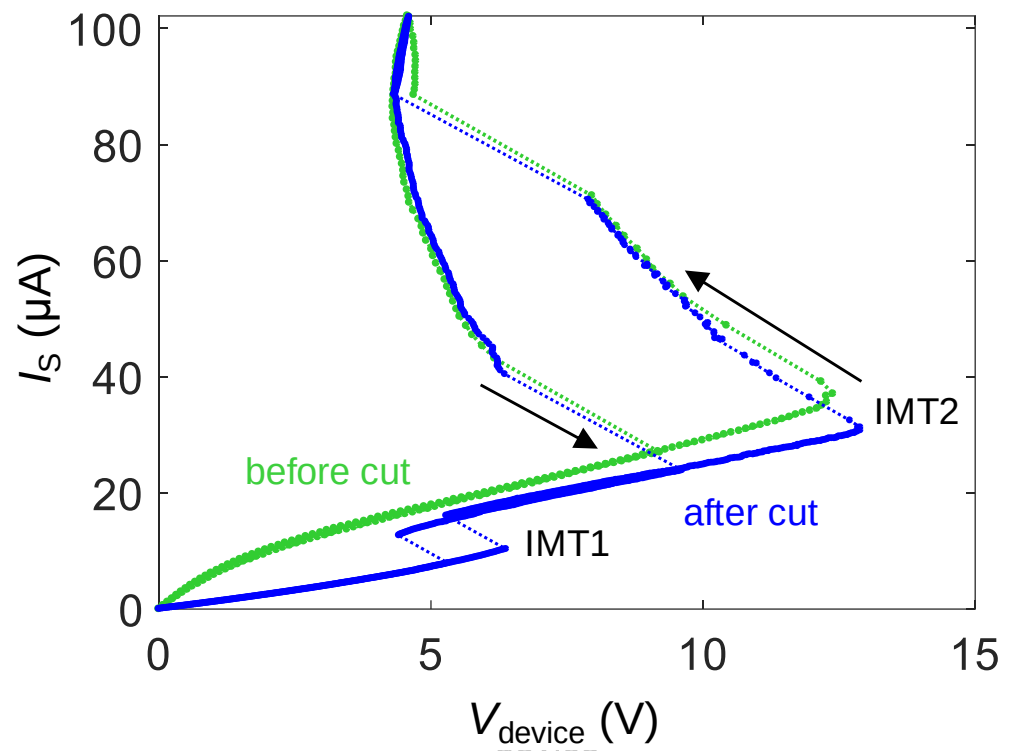

**Figure S2.** DC voltage-source switching of the nanogap device in the main text (Figures 1b-c with $L$ = 3.4 μm, $W$ = 6.3 μm) before (green) and after (blue) the CNT is cut using an AFM tip.

DC voltage-source measurements of additional gap devices are shown in Figure S3, corresponding to the devices shown in Figure S1. $V_{device}$ refers to the voltage across the Pd contacts to the CNT electrodes and is given by $V_{device} = V_S - I_S R_S$, subtracting the voltage drop across the series resistor. Because there is a significant resistance to the CNT (which is several μm long) in series with the gap, the voltage drop across the $VO_2$ within the gap will be smaller than $V_{device}$. Switching in all nanogap devices occurs in two steps, with the gap undergoing IMT (IMT1), then the connected CNT heater triggering an IMT along its length (IMT2). IMT2 occurs at nearly the same voltage as before cutting the CNT.

The on/off ratio of IMT1 (the change in the current just before and after IMT1) is typically only 1.1 – 1.6. When the gap is insulating, nearly half of the total current comes from the parallel leakage current through the $VO_2$ (far from the gap, between the Pd contacts), which is several μm wide. Once the gap is metallic, the total current is limited by the resistance of the long CNT electrodes. The total current after IMT1 is slightly lower than before cutting the CNT, due to the added interfacial resistance between the CNT ends and the $VO_2$ underneath, and because metallic $VO_2$ is slightly more resistive than an equivalent length of metallic CNT. The on/off ratio could thus be increased by making the $VO_2$ narrower around the CNT (here limited by optical lithography and the wet etching process), and the total CNT length shorter.

The IMT1 voltage can depend slightly on the bias direction and polarity, as shown in Figure S4. This may reflect some asymmetry in the current flow pathway, or in the interfaces between the CNT ends and the $VO_2$ (either in their heating, or their contact resistance). The asymmetry appears to be stronger in devices that have a bent CNT end (Figure S4a-b) compared to a gap device with a straight cut (Figure S4c), though the higher switching voltage does not always correspond to biasing the side of the CNT that was bent/dragged.

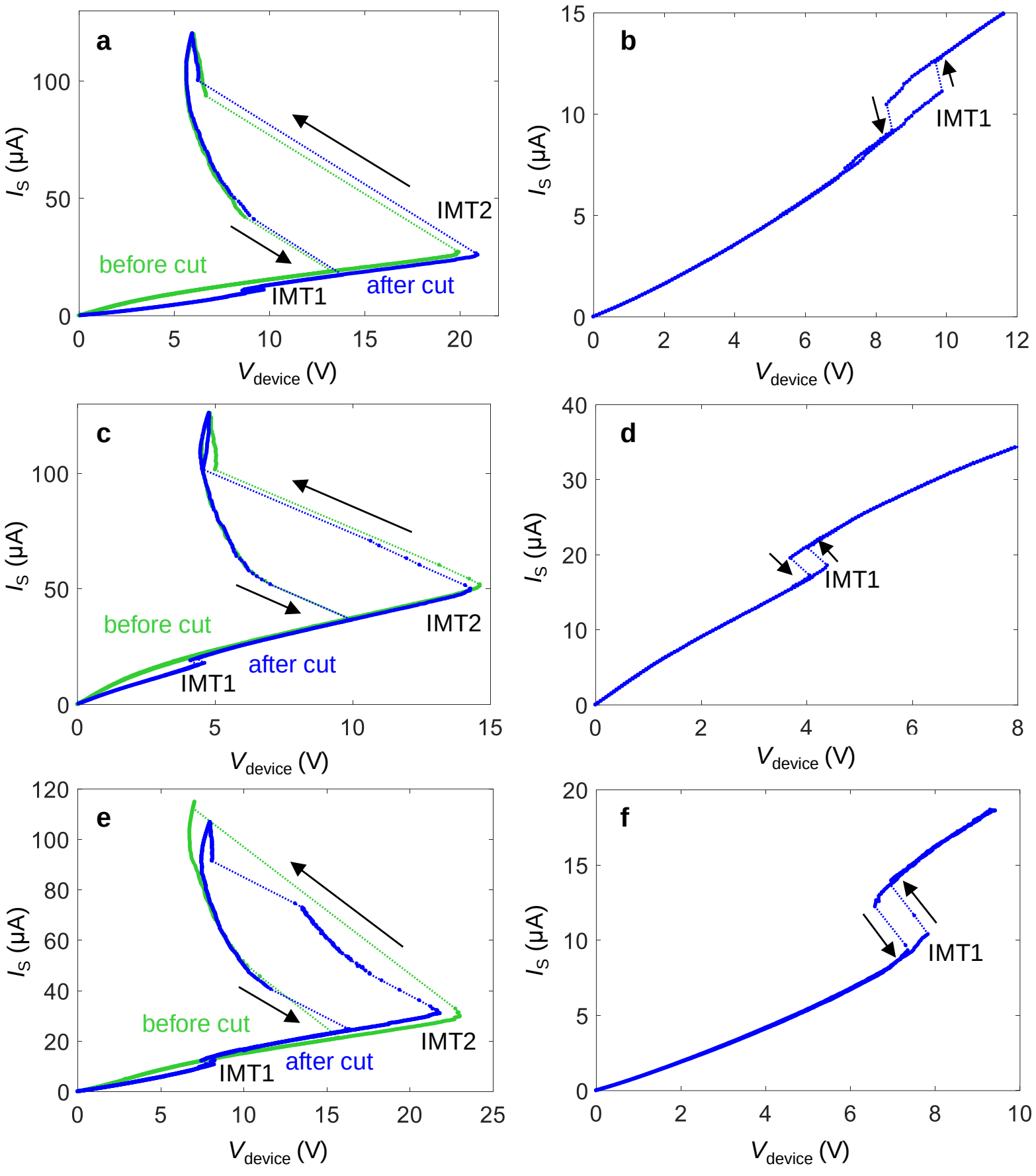

**Figure S3.** (a)-(b) DC voltage-source switching of the nanogap device shown in Figure S1a. Cutting the CNT adds IMT1 associated with the IMT of $VO_2$ within the gap of the cut CNT. (c)-(d) DC voltage-controlled switching of the gap device shown in Figure S1b. (e)-(f) DC voltage-controlled switching of the gap devices shown in Figure S1c.

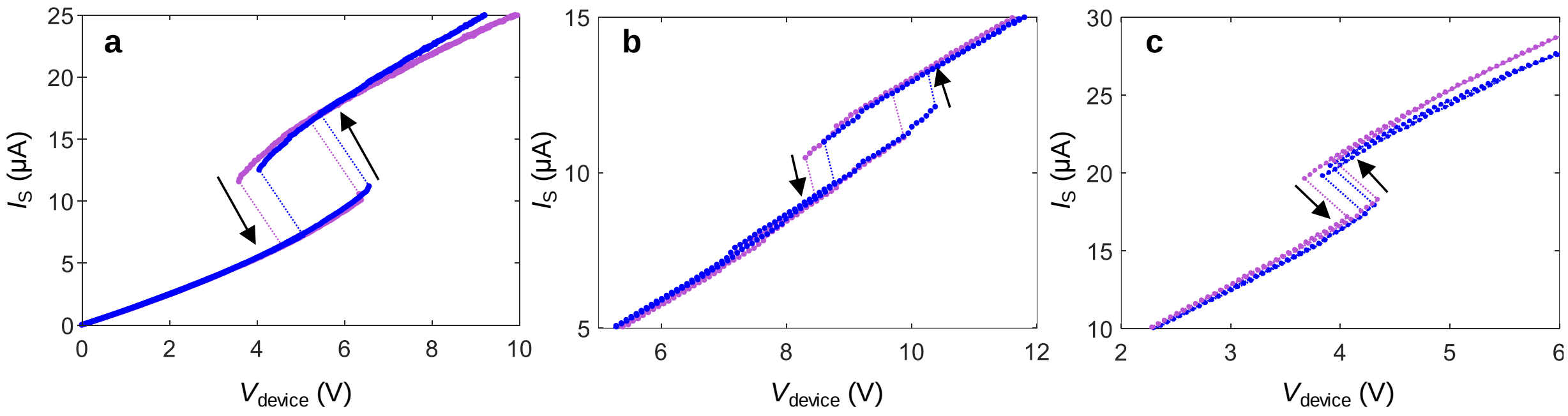

**Figure S4.** DC voltage-source measurements of IMT1 in nanogap devices with different bias directions. Relative to the AFM scans in Figure S1, a positive bias is applied to the top half of the CNT (purple) or the bottom half (blue). (a) The device in main text Figure 1 has a slightly higher switching voltage when biased on the bent side. (b) The device in Figure S1a has a slightly higher switching voltage when the straight side is biased. (c) Switching behavior of the device in Figure S1b has only very slight directional dependence.

## b. Oscillations

Oscillations in device current were measured using an Agilent InfiniiVision MSO710A oscilloscope in series with the device by measuring the voltage across its 50 Ω input impedance. The noise level of the oscilloscope is around a 1-2 mV, limiting sensitivity to changes in current close to ~µA range (in addition to background noise, the analog-to-digital converter within the oscilloscope limits output to discrete levels with a maximum resolution of 8-12 levels per 2 mV). Typical measured noise is shown in Figure S5a. Any offset from 0 V is corrected. This level of noise is observed in measurements of the nanogap device across both IMT1 and IMT2, as well as in $VO_2$ devices without any CNT. Despite the appearance of different noise levels in main text Figure 3, which is due to differences in the total voltage range (a higher range limits resolution) and time scales (different sampling density) used, a similar level of noise is observed in all devices. This noise appears to originate from the background/setup and oscilloscope limitations, and not the device itself.

Raw data without any smoothing or filtering is shown in Figure S5b for the device shown in the main text (Figures 1 and 3b). Oscillations can be seen more clearly by smoothing the data (red), such as by taking a moving average, here over a window of 20 data points (the spacing between each data point is 1 µs). For comparison, a smoothed noise waveform is shown in red in Figure S5a, showing no oscillations. The oscilloscope also has a built-in "high resolution" mode which behaves similarly to the post-processing smoothing applied in Figure S5, in which extra sampled data points are averaged over.

Data can be filtered in hardware during measurements by using the built-in bandwidth limiter on the oscilloscope, which attenuates high frequency noise (approximately >20 MHz). The result of this are shown in Figure 3b and Figure S5c, making the oscillations much clearer. A smoothed waveform calculated for this data is also shown in red. Oscillations shown in the main text for the nanogap device (Figure 3b) were collected simultaneously using the bandwidth limiter and the high resolution mode on the oscilloscope.

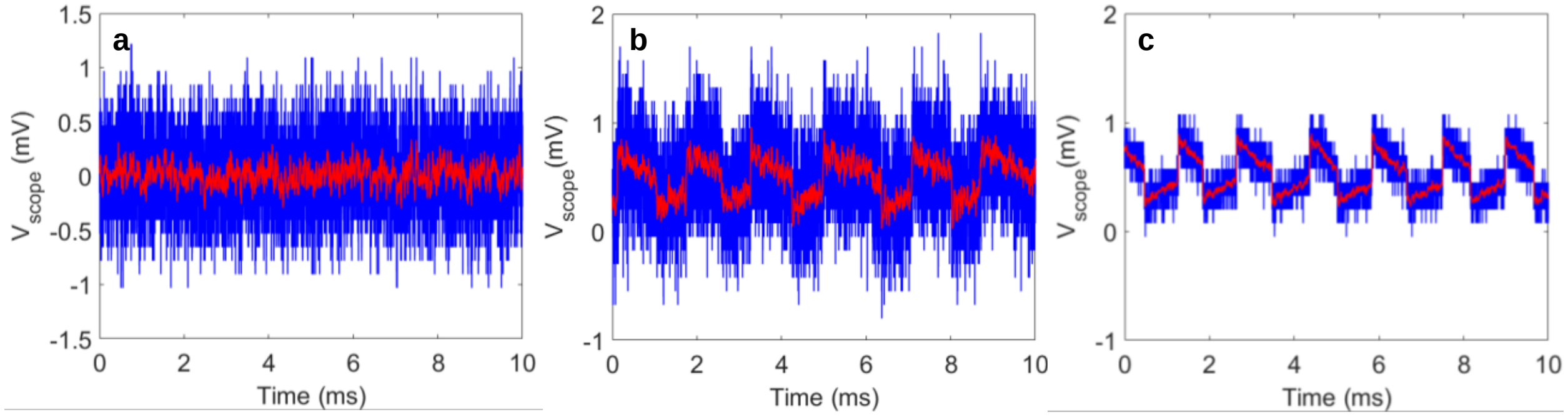

**Figure S5.** (a) Typical noise seen by the oscilloscope (blue). A smoothed waveform (red) is generated by taking a moving average. (b) Oscillations corresponding to main text Figure 3b, without any oscilloscope filtering applied (blue). A smoothed waveform (red) is made by taking a moving average. (c) Oscillations in the same device, measured with the bandwidth limiter on the oscilloscope used (blue). This significantly reduces high frequency noise. A calculated smoothed waveform is shown in red.

The energy per pulse delivered to the 50 Ω oscilloscope load is ~3 pJ/pulse and ~0.9 pJ/pulse for oscillations across IMT1 and IMT2, respectively (main text Figures 3b,c). Despite the large differences in pulse duration and magnitude, the total energy delivered to the load for the two types of oscillations is similar. Note that the energy dissipated at the device itself will be different than at the load. Both values are nearly 100 times smaller than the energy delivered to the load per pulse of a similar $VO_2$ device without any CNT heater or electrode.

We observed oscillations in several other nanogap devices during IMT1 at a similar slow ~kHz frequency, shown in Figure S6. Figure S6b corresponds to the device in Figure S1b, which had an even smaller gap size than the one shown in the main text. However, the on/off ratios of IMT1 in these devices are smaller than the one presented in the main text, so the amplitude of oscillations and the signal-to-noise ratio are both lower.

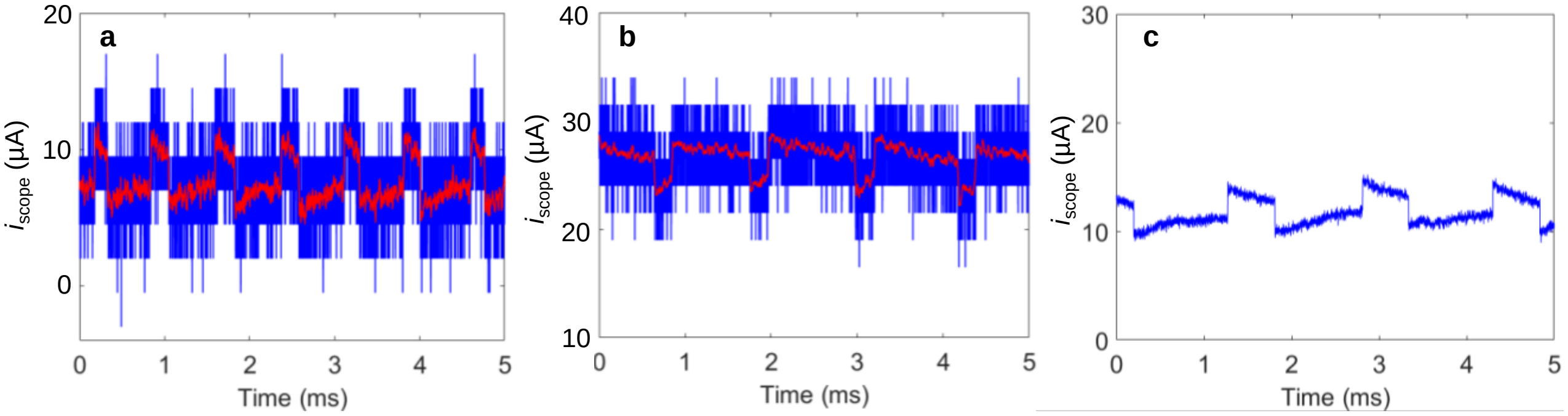

**Figure S6.** Oscillations using the bandwidth limiter of the oscilloscope (blue), with a smoothed waveform shown in red, for the nanogap devices (a) shown in Figures S1a and S3b, at $T_0$ = 295 K, (b) shown in Figures S1b and S3d, at $T_0$ = 284 K, and (c) shown in Figures S1c and S3f, at $T_0$ = 296 K. The waveform in panel c also uses the high-resolution mode of the oscilloscope.

### c. Characterization of the Impact of the Current Source on Oscillations

We note that the choice of source and settings used to apply a fixed bias current can influence, or potentially limit, the oscillation frequency of IMT devices. Figures S7a and S7d show that the oscillation frequency of $VO_2$ devices without CNT electrodes/heaters varied by nearly an order of magnitude depending on whether a Keithley 4200-SCS parameter analyzer or a Keithley 2450 SourceMeter was used to apply a 17 µA bias current to the same device and setup (including series resistor, cables, probes, etc.). Furthermore, for a given current source, the oscillation frequency decreased by 2-3× with each decade of increase in the source's output current range, even though the same (17 µA) bias was applied, shown in Figures S7a-c. We observed a similar behavior in nanogap devices during the slow oscillations of IMT1 (Figure S8). All oscillation results presented in the rest of the paper were obtained using the lowest possible current range for a given bias current, which produced the fastest oscillations.

On the one hand, an IMT event causes the device resistance to abruptly change, and therefore both the total voltage and load seen by the source change rapidly. Most commercially available parameter analyzers and current sources contain active feedback to maintain current levels, and it is possible that the source tries to respond to the changing load conditions, influencing the oscillation behavior. On the other hand, the source may contain parasitics, namely output capacitance(s), that could contribute to the dynamical behavior of the oscillator. The change in current source range may thus reflect a change in this active feedback circuit or in the parasitics, and therefore could influence or limit the maximum possible oscillation frequency in these devices. We note that the implementation of the current source circuit can therefore be an important consideration for (and a means of control over) oscillatory devices for practical applications.

In our experiments, we observed that the fast oscillations of continuous CNT heater devices [2] and of IMT2 in nanogap devices (Figure S9) were insensitive to the choice of current source and the current source range. It may be that the oscillations are fast enough, or the resistance/voltage changes small enough, to be outside the response bandwidth of the current source feedback. Or it may be that the heating term modifies the dynamics such that the device preferentially oscillates independently, driven by its own smaller thermal and electrical time constants rather than being influenced or limited by external parasitics.

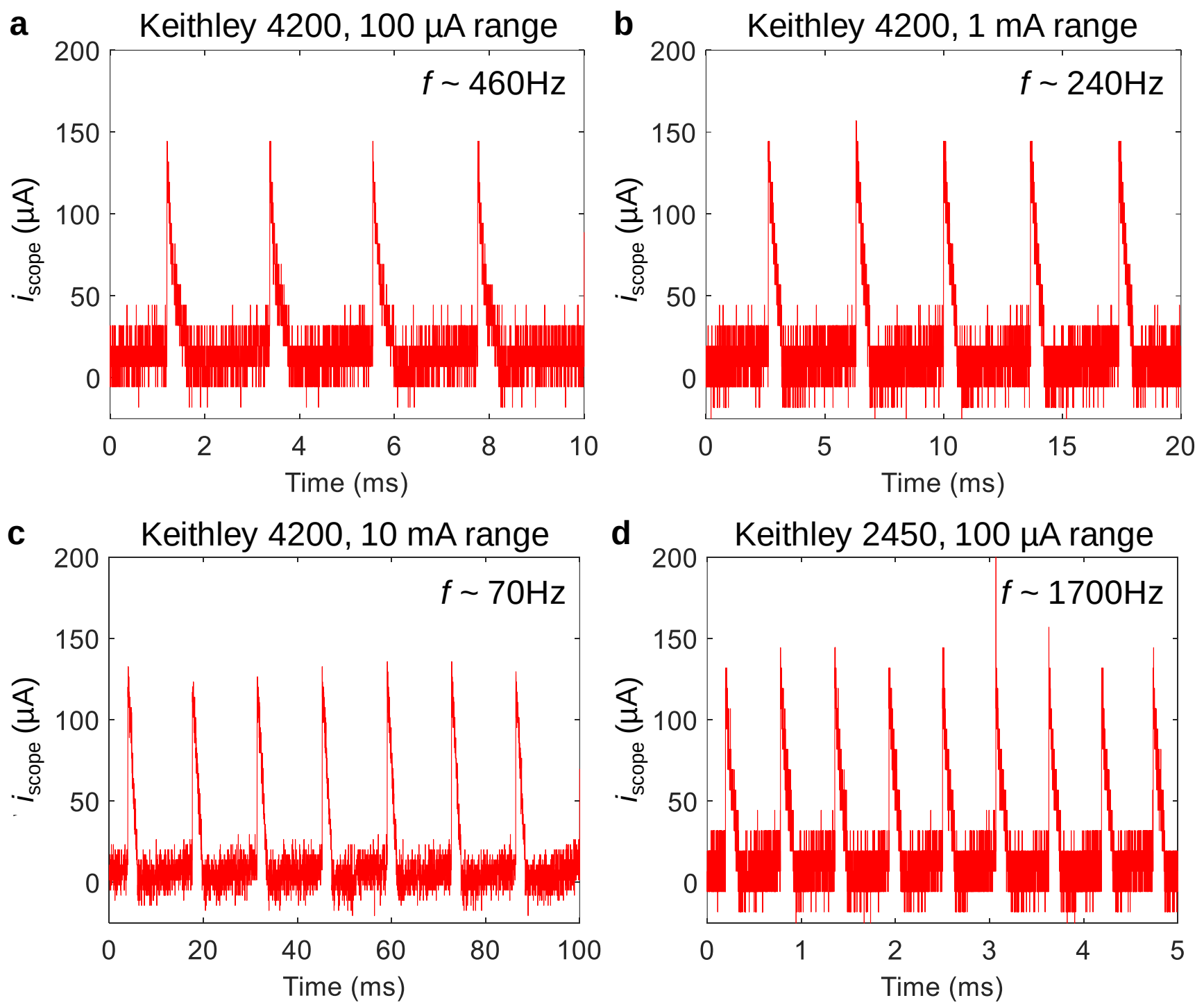

**Figure S7.** Oscillations of a micrometer-scale $VO_2$ device ($L$ = 3.5 µm, $W$ = 2.8 µm) held at a fixed current bias of $I_S$ = 17 µA applied using different current range settings on either (a)-(c) a Keithley 4200-SCS or (d) a Keithley 2450 SourceMeter. Note the very different frequencies and timescales.

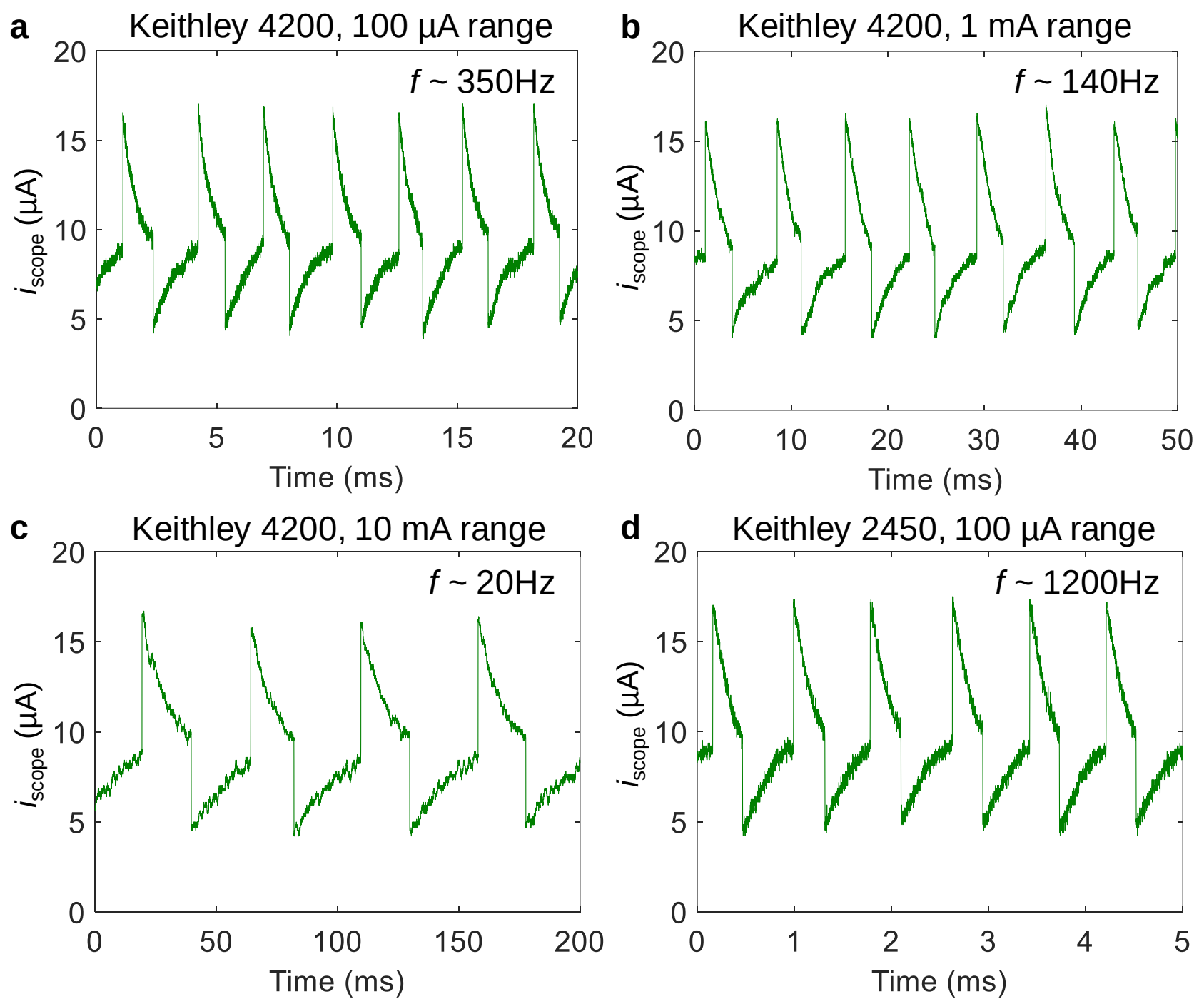

**Figure S8.** Oscillations of a nanogap device held within its IMT1 at a fixed current bias of $I_S$ = 9 µA applied using different current range settings on either (a)-(c) a Keithley 4200-SCS or (d) a Keithley 2450 SourceMeter. Similar to the larger $VO_2$ devices, the current source settings resulted in very different frequencies and timescales.

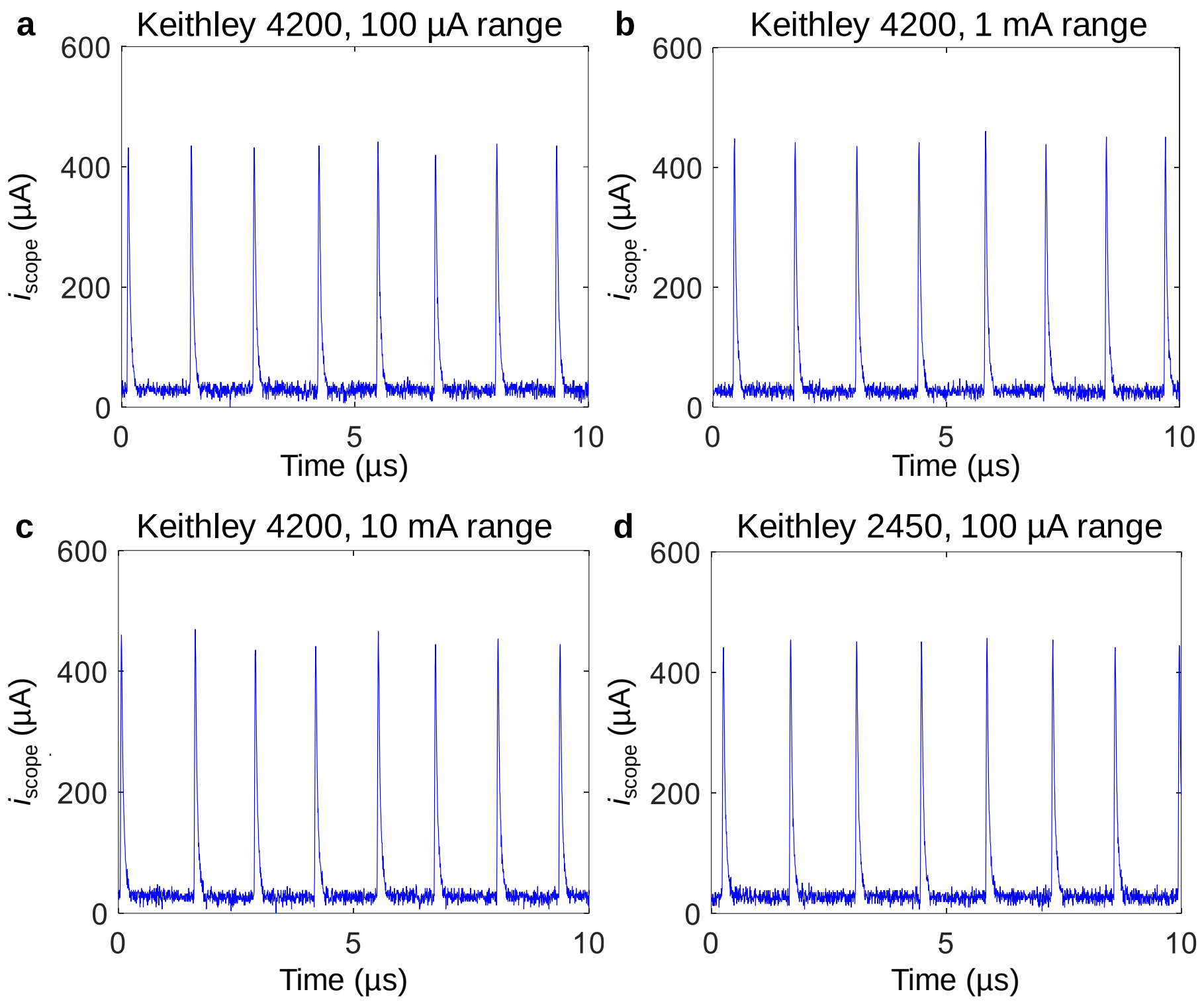

**Figure S9.** Oscillations of a nanogap device held within its IMT2 at a fixed current bias $I_S$ = 50 µA applied using different current range settings on either (a)-(c) a Keithley 4200-SCS or (d) a Keithley 2450 SourceMeter. Unlike oscillations in IMT1, the source settings have no discernable impact on frequency in IMT2 (all 0.7 MHz).

## d. Temperature Dependence

The switching voltages for both IMT1 and IMT2 decrease with increasing ambient temperature, as shown in Figure S10 for the device in main text Figure 1. This is consistent with a thermally-driven mechanism in which less power is required to heat the $VO_2$ to $T_{IMT}$ if the device is held at a higher ambient temperature. At all temperatures, the ratio of $V_{IMT1}$ to $V_{IMT2}$ remains approximately 0.55.

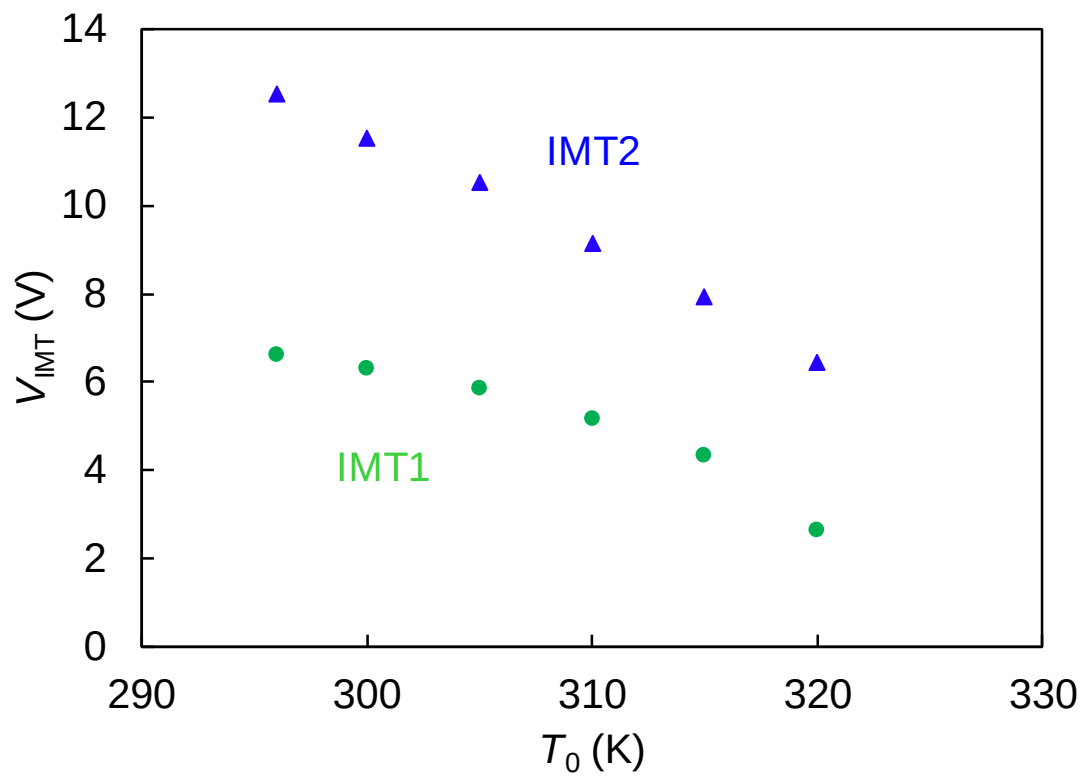

**Figure S10.** Dependence of the DC switching voltage on ambient (stage) temperature, for both IMT1 in the gap (green circles) and IMT2 along the CNT (blue triangles). Both IMT voltages decrease with increasing ambient temperature.

## e. Incubation Time and Switching Energy

A device's incubation time (switching delay) can be estimated by applying a voltage pulse and measuring the time taken to observe a jump in current at the IMT. Typically the incubation time decreases as the amplitude of the applied voltage pulse increases [8], and is associated with the time needed to heat the device to its IMT temperature (i.e., a thermal time constant), and/or with the time needed to charge capacitances (intrinsic or parasitic) and reach the switching voltage at the device terminals (i.e., an electrical time constant). The incubation times may give some insight into the switching speed of devices, but do not represent the same conditions used to generate oscillations, and do not capture the full dynamical behavior.

Because our devices had higher switching voltages than most fast voltage pulse sources can supply (typically limited to 10 V amplitude), the pulsed units in a Keithley 4200-SCS parameter analyzer were used to generate higher voltage pulses up to 40 V. The rise and fall times of pulses were limited by the tool to 100 ns. In nanogap devices with a lower switching voltage, an Agilent 81150A pulse generator was used, which had a faster rise time. However, devices exhibited ringing below ~70 ns rise times, so the results with 100 ns rise times are shown in Figure S11 for better comparison with other devices. The voltage across a 50 Ω oscilloscope in series with the device was used to measure the device's current waveform. Example waveforms for a $VO_2$ device without a CNT, a $VO_2$ device with a continuous (uncut) CNT heater, and a $VO_2$ nanogap device with a cut CNT are shown in Figures S11a-c, respectively.

Figure S11d-f shows extracted incubation times for these devices, with the left edge of each plot (lower limit of the horizontal axis) set as the static switching voltage for the device. The time for switching was taken as the start of the pulse (including rise time) to the start of the rising edge of the capacitive overshoot spike produced upon IMT. At high voltages, the extracted incubation time is thus limited by the rise time of the pulses (0.1 μs). The incubation times of all devices decreased with increasing pulse amplitude. However, note that the pulse voltage at the source is plotted, which is not the same as the voltage at the Pd contacts or at the nanogap (due to parasitics, contact resistance and the series resistance of the long CNT electrodes). All three devices could achieve similar timescales for incubation, though the CNT device was slightly slower because of its longer length.

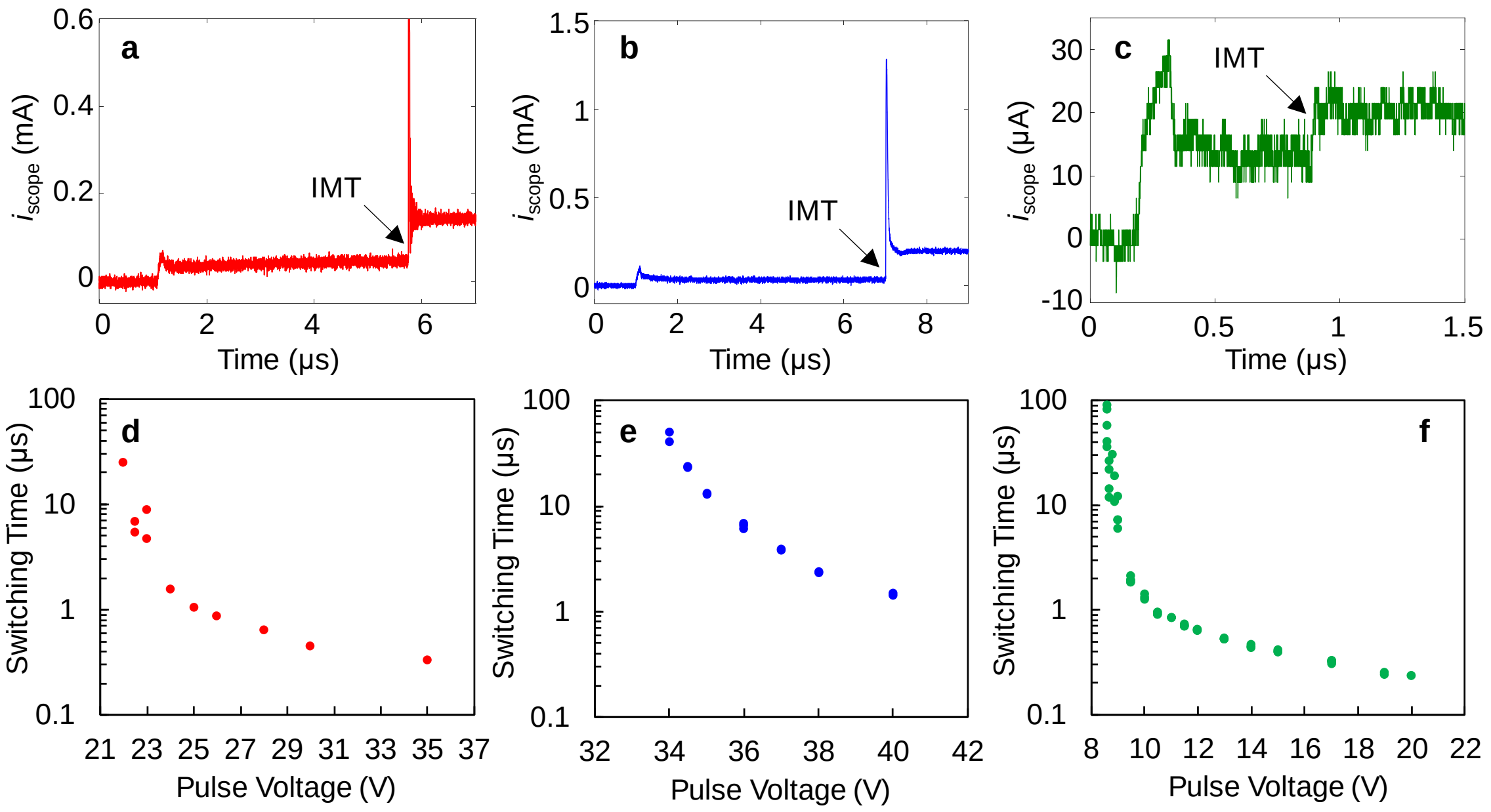

**Figure S11**. Current waveforms when (a) a 23 V voltage pulse is applied to a $VO_2$ device without a CNT ($L$ = 1 μm, $W$ = 5 μm), (b) a 36 V voltage pulse is applied to a $VO_2$ device without a CNT ($L$ = 7 μm, $W$ = 6 μm), and (c) an 11.5 V voltage pulse is applied to a $VO_2$ gap device with CNT electrodes ($L_{gap}$ < 100 nm, $L$ = 3.4 μm, $W$ = 6.3 μm). (d)-(f) Incubation times at different pulse amplitudes for the same devices.

Switching energy (Figure S12) could be estimated by integrating device power over time using the measured waveforms. The voltage across the device terminals was not directly measured because the 1 MΩ input impedance of an oscilloscope in parallel with the device will disturb its operation if used to measure the device's voltage, given that the insulating state resistance of a $VO_2$ device is on the order of a few MΩ. Thus, the device voltage was approximated by subtracting the voltage drop across the series resistor from the input voltage pulse waveform (using $V_{device} \sim V_S - I_{device}(t)R_S$). In the presence of parasitic capacitors, the voltage seen by the device may be even lower, and thus the switching energy estimates presented here are upper bounds. While the $VO_2$ device without a CNT and the $VO_2$ device with a continuous CNT heater showed similar switching energies on the order of nJ, the nanogap device had a much lower switching energy down to 85 pJ. This is consistent with smaller thermal and switching volumes in the nanogap device.

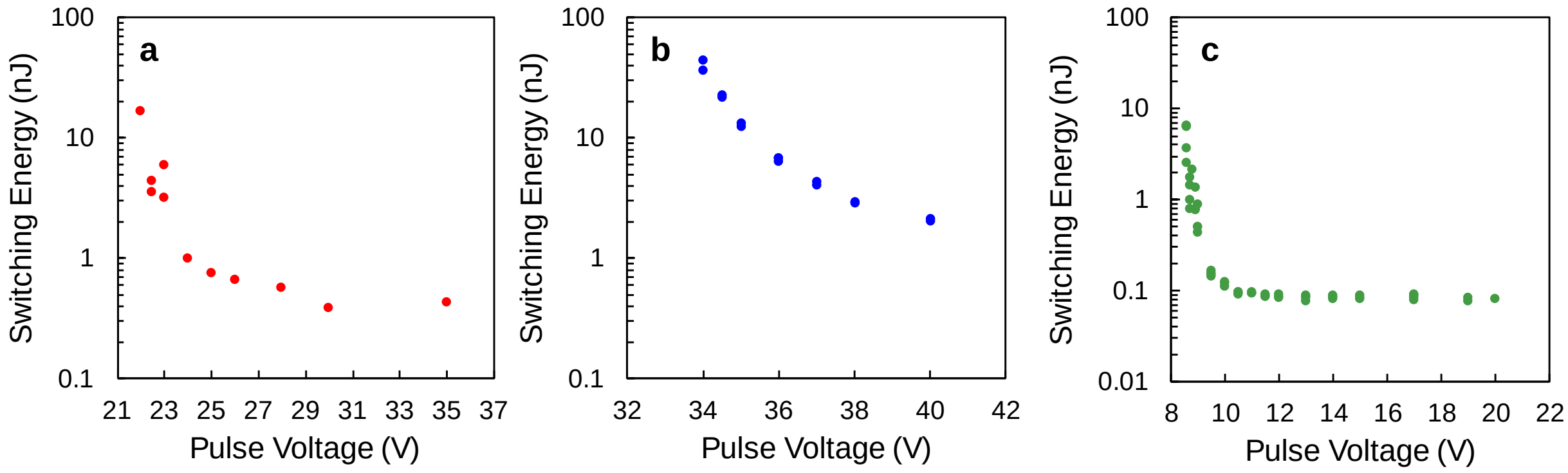

**Figure S12.** Estimated upper bounds of switching energies for (a) a $VO_2$ device without a CNT ($L$ = 1 μm, $W$ = 5 μm), (b) a $VO_2$ device with a continuous CNT heater ($L$ = 7 μm, $W$ = 6 μm), and (c) a $VO_2$ nanogap device with CNT electrodes ($L_{gap}$ < 100 nm, $L$ = 3.4 μm, $W$ = 6.3 μm).

## 3. Additional Scanning Probe Microscopy Measurements

### a. Kelvin Probe Microscopy

Kelvin Probe Microscopy (KPM) was done in dual pass mode while a constant external voltage bias was applied to the device in Figure S1a and S3a. One of the Pd electrodes was grounded, and the other was connected to the 200 kΩ series resistor to which the positive bias $V_S$ was applied. These were outside the top and bottom of the scan area, respectively. During each scan, the device current remained steady and set in its insulating or metallic state. A full series of KPM images at different biases is shown in Figure S13, after a first-order flattening operation was applied.

Before flattening, the surface potential along the device from grounded to positive electrode shows a nearly linear increase, corresponding to a constant electric field. However, because the $VO_2$ in the nanogap is more resistive than the CNT there is a higher field across it, corresponding to a sharper slope (Figure S14a). Flattening removes the average slope of the potential and highlights any local changes in field across the device, namely within the nanogap. There is a much larger contrast observed in flattened images of the insulating state over the metallic state because of the larger deviation in field across the $VO_2$ in the gap relative to the metallic CNT, as shown in Figure S14b.

Reversing the polarity or direction of applied bias reverses the field direction, and therefore inverts the contrast seen in the flattened image, but does not otherwise affect the surface potential profile or device operation (Figure S15).

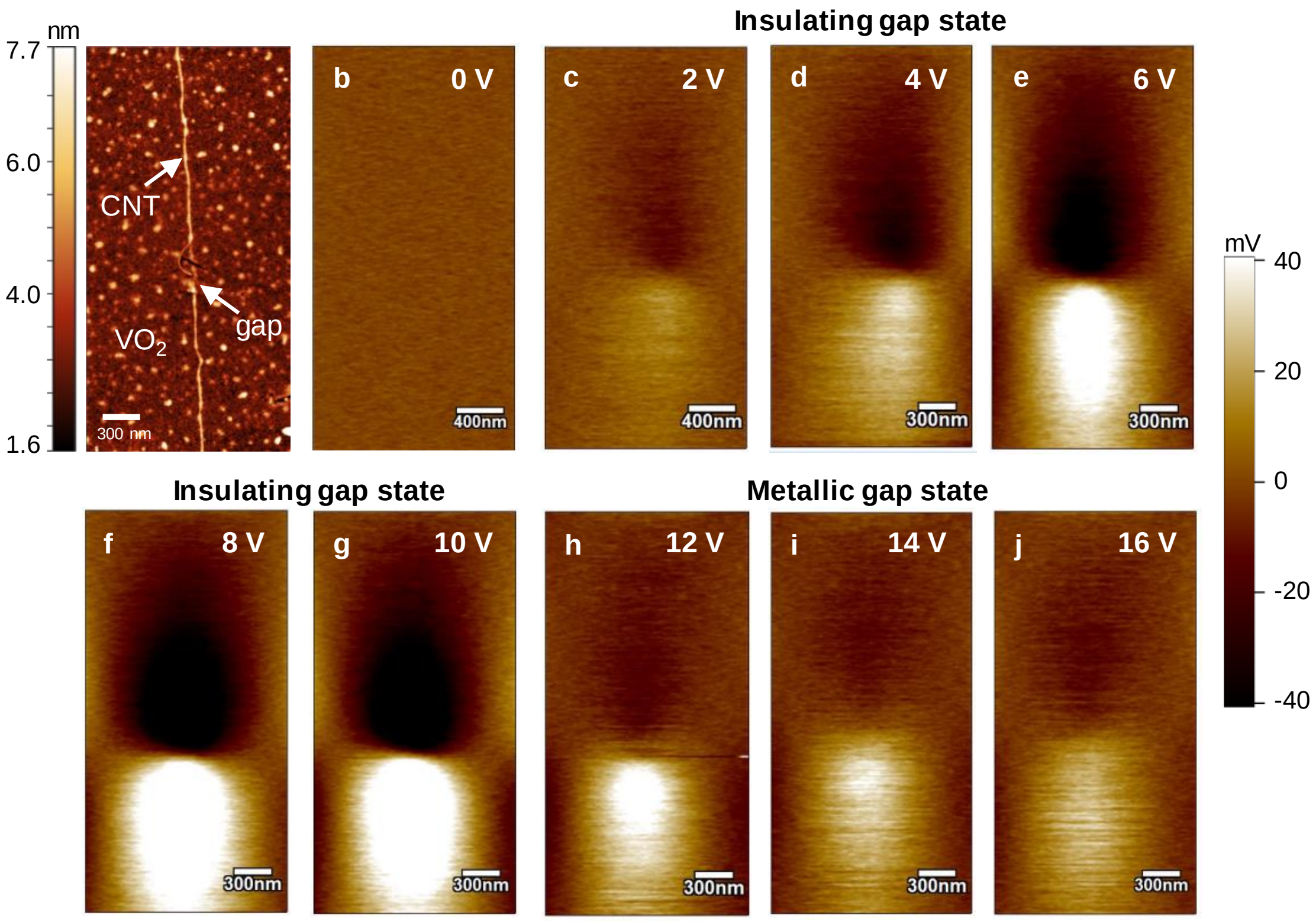

**Figure S13.** Full set of KPM images of a nanogap device. (a) Topography of the cut CNT in a nanogap device, with an arrow indicating the gap location. (b) KPM of the same device and location with no device bias. (c)-(j) Flattened KPM images of the device with increasing bias. The Pd electrodes are outside the top and bottom margin of the images, with the bottom electrode biased positive. Images (c)-(g) are of the insulating $VO_2$ state in the nanogap, and (h)-(j) are of the metallic nanogap state. For better comparison, the same scalebar for surface potential is used in all images (some images are saturated).

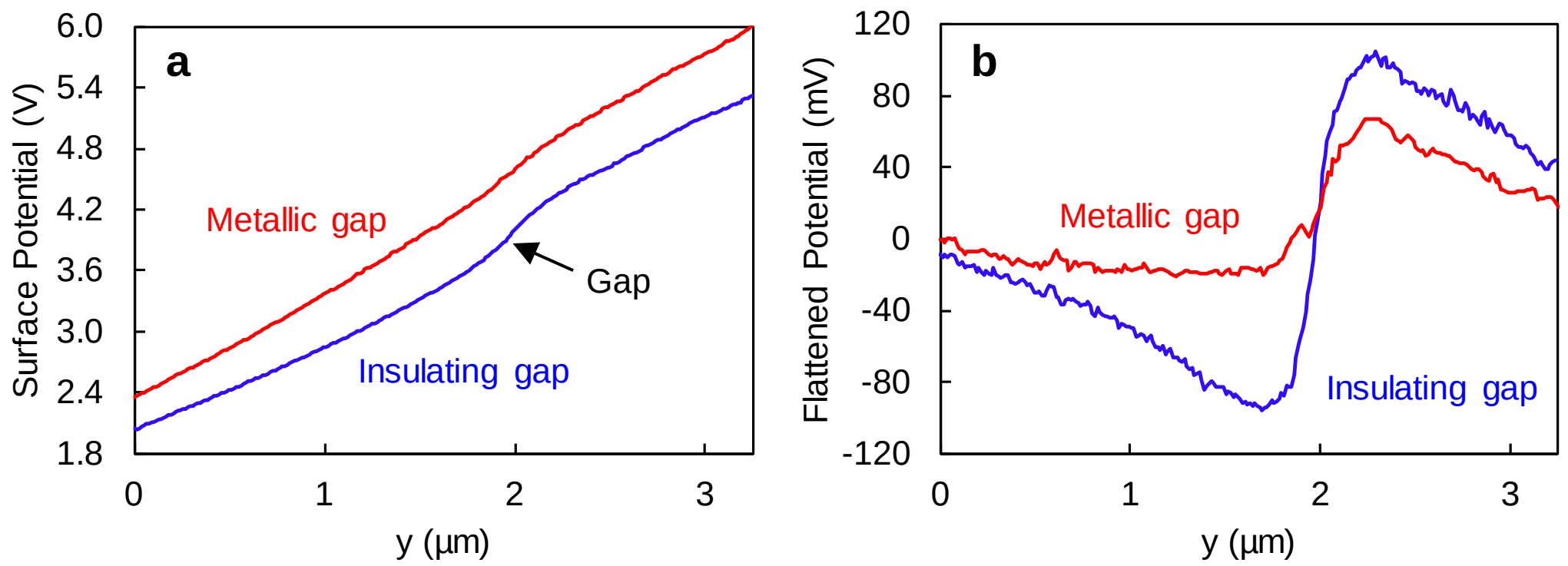

**Figure S14.** (a) KPM surface potential profiles along the CNT prior to flattening, for a 10 V applied bias (red) with an insulating gap state and 12 V applied bias (blue) with a metallic gap state. (b) KPM surface potential profiles along the CNT after a first-order flattening operation to effectively remove the average slope of each line in (a). These correspond to profiles along the vertical direction in Figure S13g-h.

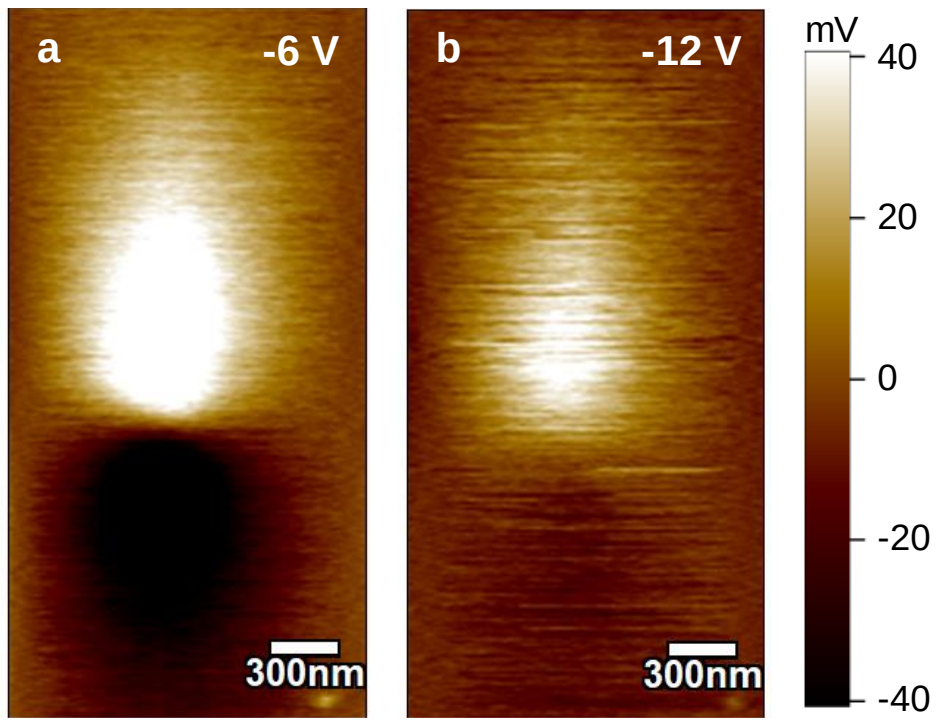

**Figure S15**. KPM of the nanogap device in Figure S13 with the bias polarity reversed (a negative bias is applied to the series resistor and contact beyond the top of the scan, and the contact outside the bottom of the scan is grounded), showing (a) the insulating gap state and (b) the metallic gap state.

## b. Scanning Thermal Microscopy

Scanning thermal microscopy was performed in passive mode with a contact mode setpoint of 0.5 V and a 0.5 V tip bias. The SThM tip is made of Pd on SiN (model PR-EX-GLA-5 from Anasys®), with a tip radius of <100 nm, that senses changes in device heating via a change in tip resistance. The tip requires electrical isolation from the biased device, and thus devices were capped with a 35 nm layer of 2% 495K poly(methyl methacrylate) (PMMA) in anisole. The PMMA was spin coated at 7000 rpm for 40 s, followed by baking on a hotplate at 180°C for 5 min. A PMMA layer was used because oxide deposition can reduce the stability of the CNT and $VO_2$. All scans were taken on the surface of this PMMA layer, and thus the thermal profile observed is expected to be broader than that of the $VO_2$ surface.

A set of SThM images at multiple $V_S$ bias points is shown in Figure S16. The same nanogap device was used as for the KPM measurements, corresponding to Figures S1a and S3a. A 200kΩ series resistor was used for all measurements. Because $V_S$ = 10 V was very close to IMT1, and the threshold voltage was prone to shifting very slightly over time and cycles, this bias point corresponds to the insulating state in KPM but the metallic state in SThM.

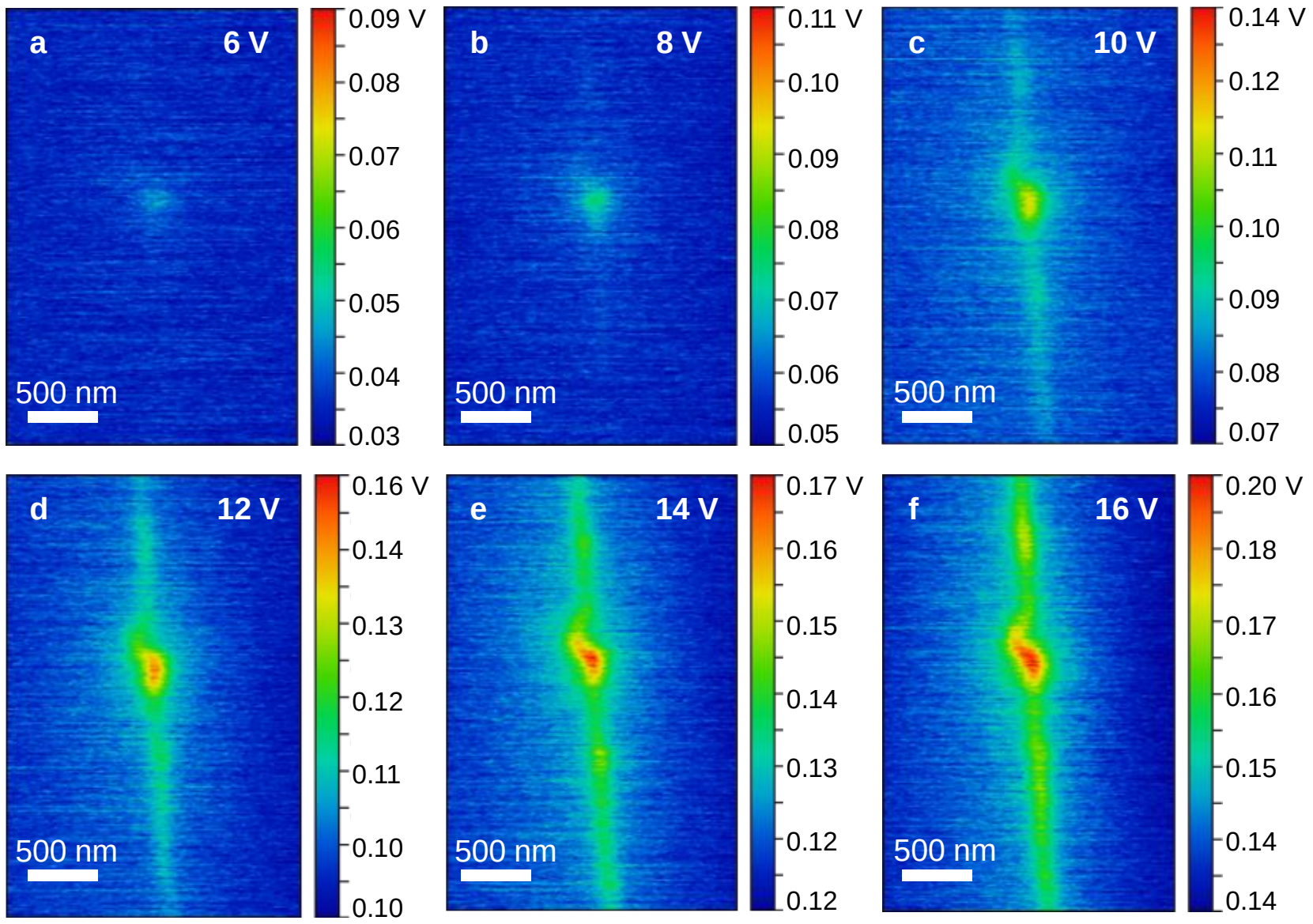

**Figure S16.** SThM images showing heating profiles of a nanogap device at different device biases, as seen on the surface of a 35 nm PMMA capping layer. (a)-(b) When the $VO_2$ within the CNT nanogap is insulating, heating is highly localized to the gap. (c)-(f) After IMT1, the $VO_2$ in the gap is metallic and the CNT is connected, causing heating along its length.

## 4. Finite Element Modeling

We constructed a 3D electrothermal finite element model for nanogap devices using COMSOL Multiphysics. This self-consistently coupled together calculations for the voltage (*V*) and current density (*J*) in the device (Eqn. (S1), which depends on temperature $T$ through the electrical conductivity σ) with calculations for the temperature (Eqn. (S2) which are affected by Joule heating).

$$\nabla \cdot (\sigma(x, y, z, T)\nabla V) = 0 \tag{S1}$$

$$\nabla \cdot (k\nabla T) + J \cdot E = 0 \tag{S2}$$

$E$ is the electric field in the device, and $k$ is the thermal conductivity of $VO_2$. For the $VO_2$, the electrical conductivity was scaled with local temperature based on resistance measurements of $VO_2$ devices without a CNT as a function of ambient temperature. At room temperature, a $\sigma$(296 K) = 80 S/m was estimated after subtracting contact resistance, and was scaled according to experimental temperature dependence to produce a $\sigma(T)$ function. A nearest-neighbour interpolation of the function was used (with steps of ~1 K), which rounds the temperature of an element to the nearest experimental data point and uses the corresponding resistance. These discrete steps are necessary to avoid convergence issues. If a smooth curve or a linear interpolation is used within the IMT region (which is highly nonlinear) then a slight change in temperature can cause a large change in resistance from one solver step to the next, making convergence difficult.

The conductivity of the CNT is given by [9]:

$$\sigma_{\mathrm{CNT}}(T, V, L) = \frac{4q^2}{h}\frac{\lambda_{\mathrm{eff}}}{A} \tag{S3}$$

$$\text{where } \lambda_{\mathrm{eff}} = \left(\lambda_{\mathrm{AC}}^{-1} + \lambda_{\mathrm{OP,ems}}^{-1} + \lambda_{\mathrm{OP,abs}}^{-1} + \lambda_{\mathrm{defect}}^{-1}\right)^{-1} \tag{S4}$$

$\lambda_{\mathrm{eff}}$ is an effective electron mean free path (MFP) that combines contributions from elastic electron scattering with acoustic phonons ($\lambda_{\mathrm{AC}}$), as well as inelastic electron scattering by optical phonon absorption ($\lambda_{\mathrm{OP,abs}}$) and emission ($\lambda_{\mathrm{OP,ems}}$). Emission is influenced by the electric field, so this term is dependent on the voltage and CNT length, and all MFPs are a function of temperature. Values of $\lambda_{\mathrm{OP,300}}$ = 20 nm and $\hbar\omega_{\mathrm{OP}}$ = 0.2 eV are used here. An additional scattering term for defects ($\lambda_{\mathrm{defect}}$), with a mean free path of 1 μm, is added to better represent *I-V* characteristics of our imperfect CNTs. For additional details, see Refs. [2,9].

The geometry of the simulated devices is shown in Figure S17. For simulations shown in main text Figure 2 the actual experimental CNT shape and location were extracted from AFM images (Figure S1a) and used in the simulation (including a lateral offset in the width, and the curvature of the CNT end). For these simulations only, a 35 nm capping layer made of PMMA was included. For all other simulations, the CNT was centered in the width of the device and only half the device width was simulated due to symmetry. The nanogap was centered at $L/2$ and its length was varied. The simulated $TiO_2$ substrate was 2 μm thick (unlike the ~500 μm experimental $TiO_2$), which was sufficient to capture its thermal resistance. The electrical and thermal conductivities used for the various materials are listed in Table 1.

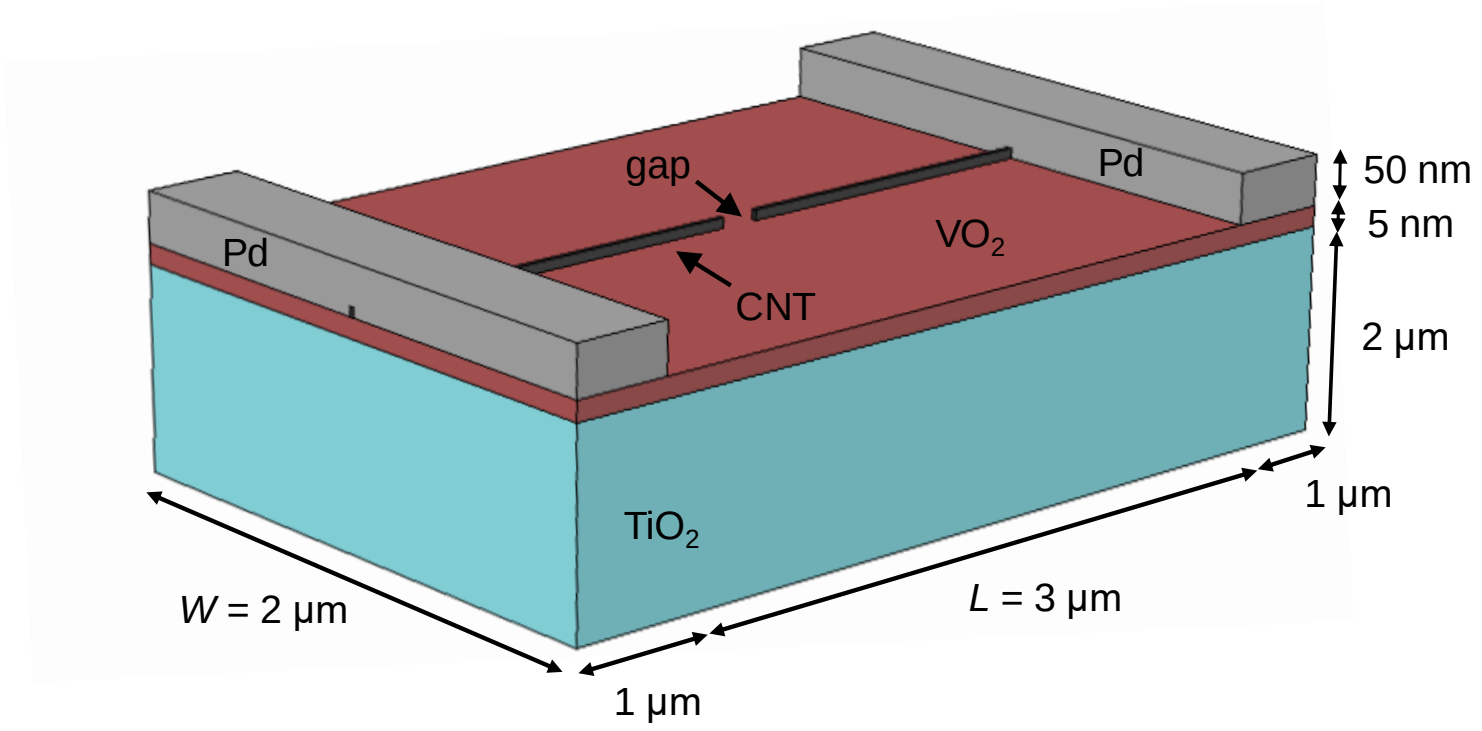

**Figure S17.** The structure used to simulate the nanogap devices with an electrothermal finite element model. Only half the width was simulated due to symmetry. For simulations in main text Figure 3, experimental CNT and $VO_2$ dimensions were used, with $L$ = 5 μm and $W$ = 5 μm (the full width was simulated).

The top of one Pd electrode was held at a fixed voltage, and the other grounded. An electrical contact resistance was included for various interfaces: CNT/Pd (25 kΩ), CNT/$VO_2$ (~100 kΩ), and $VO_2$/Pd. The $VO_2$/Pd interface resistivity was set to $1\times10^{-3}$ $\Omega\cdot cm^2$ and reduced with increasing local temperature, in the same manner as the $VO_2$ electrical resistivity. When simulating a voltage sweep, the solution for the previous voltage point was used as the initial conditions for the next voltage.

The bottom of the $TiO_2$ was set to room temperature (296 K), and all other boundaries to thermally insulating. A thermal boundary resistance was included for various interfaces: $TiO_2$/$VO_2$ ($1\times10^{-8}$ $m^2K/W$), $VO_2$/CNT ($5\times10^{-9}$ $m^2K/W$), $VO_2$/Pd ($1\times10^{-8}$ $m^2K/W$), CNT/Pd ($1\times10^{-8}$ $m^2K/W$), and PMMA/$VO_2$ ($1\times10^{-8}$ $m^2K/W$). Experimental values for these particular interfaces have not been measured, so average values within typical ranges were used [10].

**Table 1.** Material properties used in simulation

| | **σ [S/m]** | **k [$Wm^{-1}K^{-1}$]** |
|---|---|---|
| **$TiO_2$** | $10^{-7}$ | 7 |
| **$VO_2$** | $T(x,y,z)$ spanning 80 to $2\times10^6$ | 3 |
| **CNT** | Eqns. S3, S4 | 600 |
| **Pd** | $10 \times 10^6$ | 72 |
| **PMMA** | $10^{-10}$ | 0.1 |

The model used a segregated approach, solving first for temperatures and then for electric fields. Each step used the direct PARDISO solver (as opposed to an iterative solver). A damping factor of 0.2 was used in the temperature step, and 0.8 in the electric field step. Reducing the damping factor results in a lower change between solver steps and aids in convergence, by avoiding overshoots across the IMT and MIT (especially in temperature).

Simulated voltage sweeps (with increasing voltage only) are shown in Figure S18a for different nanogap sizes. The IMT1 voltage ($V_{IMT1}$) scales approximately linearly with gap size, as shown in Figure S18b, and corresponds to switching of the $VO_2$ between the cut CNT ends (and a small volume just underneath the CNT ends, as the $VO_2$/CNT contact resistance generates heat). A second IMT event (IMT2) occurs when $VO_2$ along the entire length of the CNT is heated to its transition temperature. The IMT2 switching voltage at $V_{device}$ ~ 9 V is nearly the same as a simulated device with an uncut CNT. Because the IMT2 voltage does not significantly vary with gap size but $V_{IMT1}$ does, the ratio $V_{IMT1}/V_{IMT2}$ increases with increasing gap size. If the gap size is sufficiently large (here >100 nm), then only a single IMT event occurs, at a voltage between that of a $VO_2$ device with an uncut CNT and a $VO_2$ device with no CNT.

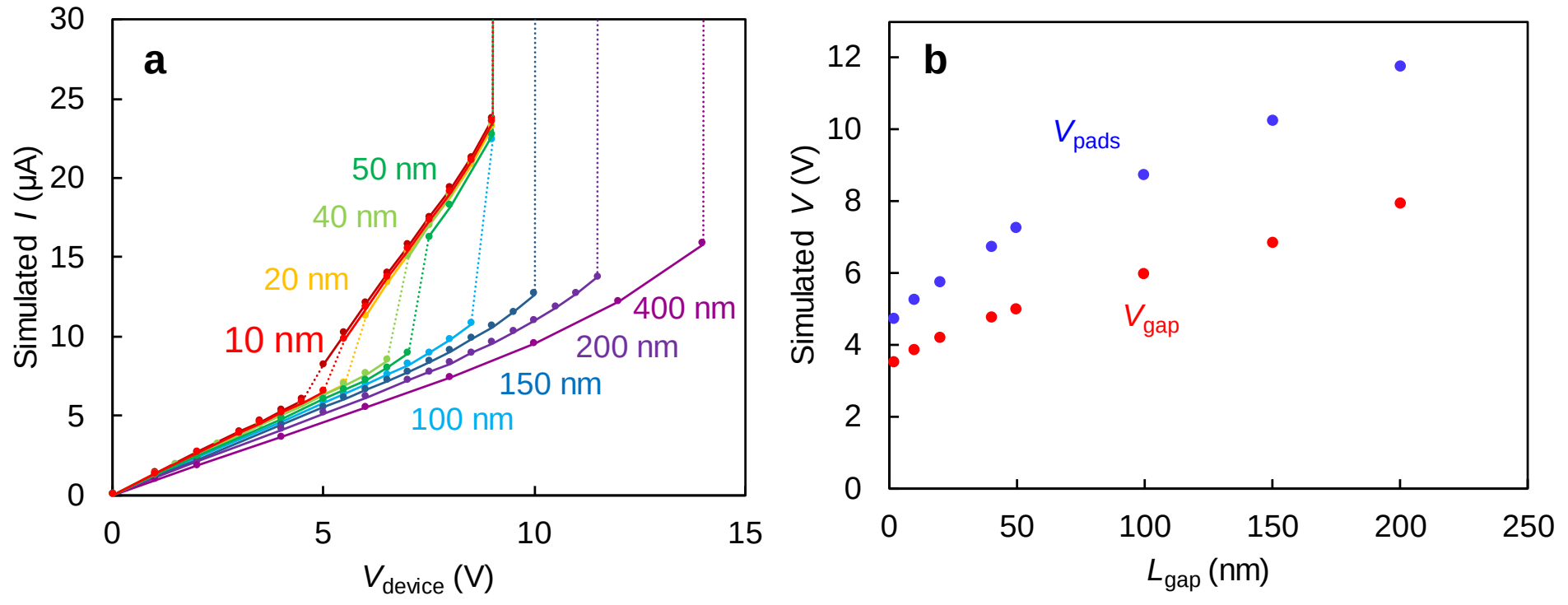

**Figure S18.** (a) Simulated *I-V* characteristics for different nanogap lengths ($L_{gap}$) in the CNT, with fixed Pd contact spacing $L$ = 3 µm. Increasing the gap size results in a higher IMT1 switching voltage and a large change in current across the IMT, but a similar IMT2 voltage. (b) IMT1 switching voltage at the Pd contacts (blue, $V_{IMT1} = V_{pads}$) and as seen at the ends of the CNT (at the contacts to the gap, in red, $V_{gap}$) with the series resistance of the CNT and the Pd contact resistance subtracted.

Figure S19 shows the dependence of simulated IMT1 and IMT2 switching voltages on ambient temperature for a 50 nm gap length (with $L$ = 3 µm). The results match the trends in Figure S10, which are consistent with a thermally-triggered IMT. The $V_{IMT1}/V_{IMT2}$ ratio is higher in simulations than experiment. In real devices the CNT quality and contacts may be worse, leading to a higher $V_{IMT2}$.

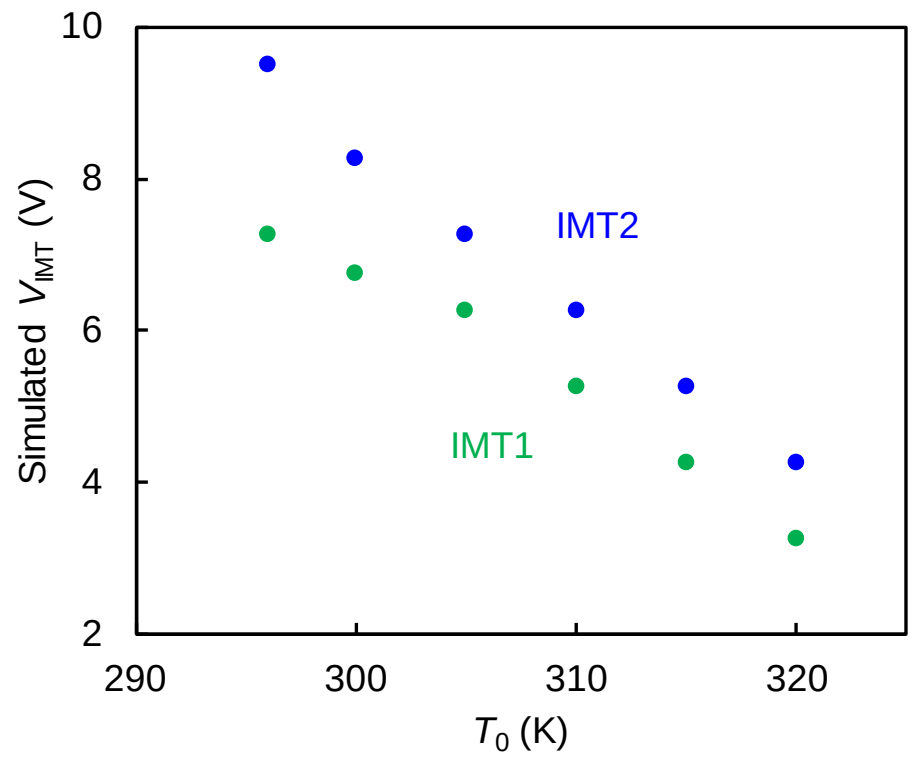

**Figure S19**. The switching voltage of both IMT events decreases with increasing ambient temperature in a device with a 50 nm CNT gap, similar to experiment in Figure S10.

## 5. Compact Modeling in LT Spice

To examine the role of various electrical and thermal parameters on the oscillation frequency of resistive switching devices, we constructed a compact model using LT Spice. For an explanation of relaxation oscillators and how oscillations can be produced, please see the Supplemental Material of [11].

In this model, the device current $i_m$ was controlled by nonlinear thermally-activated Schottky transport, given by:

$$i_m = AT^2 \exp\left(\frac{(\beta\sqrt{V_m/d} - \phi)}{k_B T}\right) \tag{S5}$$

$v_m$ is the voltage across the device, $A$ is a constant (the device cross-sectional area multiplied by the Richardson constant), and $d$ its length. $\phi$ is an energy barrier, $k_B$ is the Boltzmann constant, and $\beta$ is a constant that depends on the dielectric constant [12]. $T$ represents an average device temperature set by the thermal dynamics, which are approximated using Newton's law of cooling, as described in the main text:

$$C_{\text{th}}\frac{\mathrm{d}T}{\mathrm{d}t} = i_{\text{m}}v_{\text{m}} + \frac{v_{\text{m}}^2}{R_{heater}} - \frac{T-T_0}{R_{\text{th}}} \tag{S6}$$

$T_0$ is the ambient temperature, $C_{\text{th}}$ is the thermal capacitance of the device (which scales with the switching volume heated and cooled across the transition), and $R_{\text{th}}$ is a thermal resistance associated with the efficiency of heat transport out of the switching volume. $R_{\text{heater}}$ (if any) is a resistive heater electrically in parallel with the device, which contributes to an increase in device temperature by Joule heating. $C_{\text{th}}$ was varied between $1 \times 10^{-14}$ J/K to $5 \times 10^{-17}$ J/K, and if included, $R_{\text{heater}}$ was varied between 400 kΩ and 3 MΩ. Values of other parameters are listed in Table 2.

The device *M* was integrated into a circuit with several additional resistances and capacitances, shown in Figure S20. These include the external series resistor $R_{\text{s}}$ and the oscilloscope, which has an input impedance of 50 Ω, placed in series with the device. Several resistances and capacitances representing the device contact resistance, and device plus parasitic capacitance were also included. The circuit was solved simultaneously and self-consistently with Eqns. (S5) and (S6) in LT Spice.

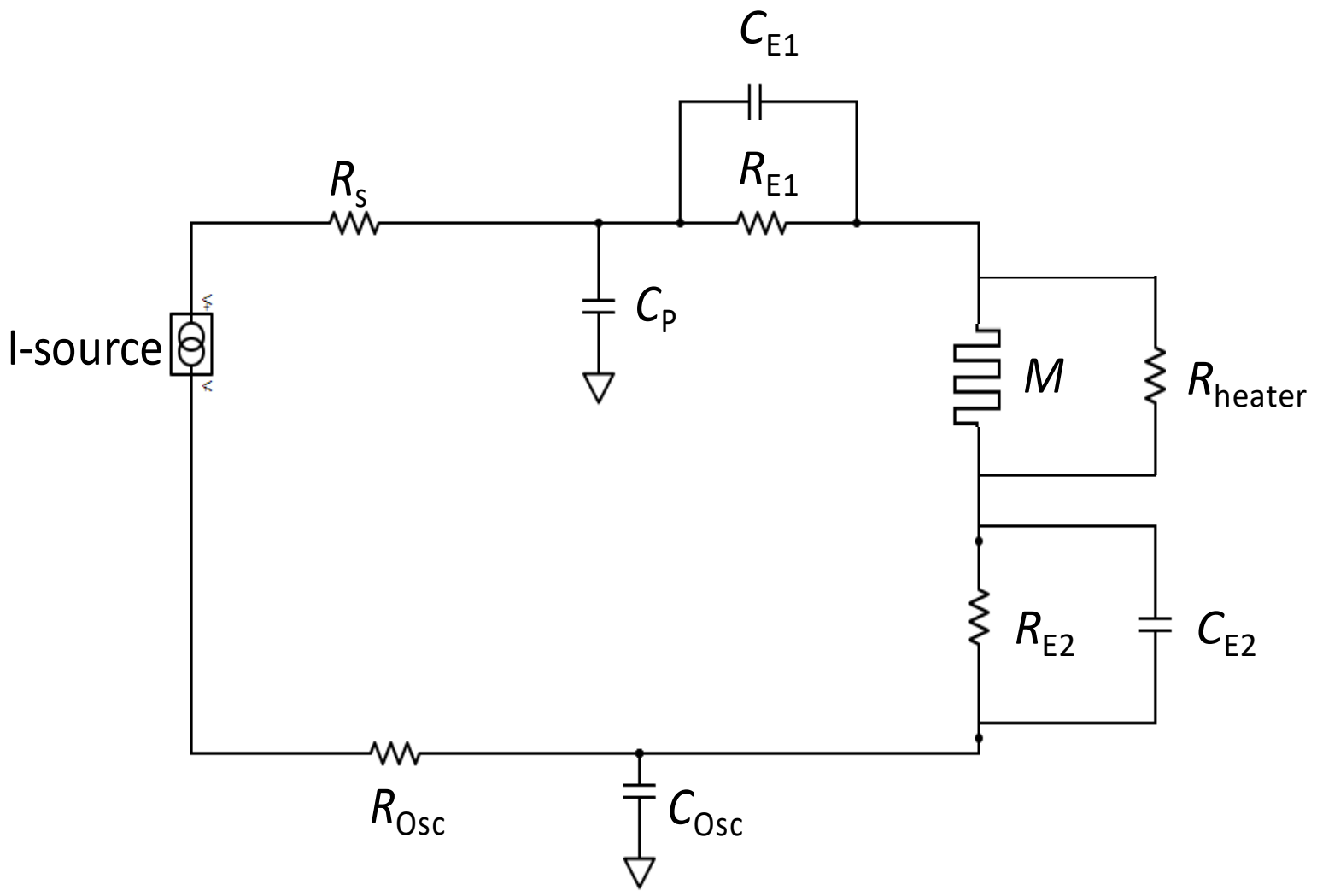

**Figure S20.** Circuit schematic used for simulations in LT Spice. The circuit consists of a switching device *M* in parallel with a thermally coupled resistive heater $R_{\text{heater}}$ (such as a CNT), an external current source, an external series resistor $R_{\text{S}}$, an oscilloscope in series represented by $R_{\text{osc}}$ and $C_{\text{osc}}$, and additional device resistances and capacitances. E1 and E2 refer to the two electrodes of a planar device).

**Table 2.** Parameters used in SPICE Simulations

| **Parameter** | **Value** |
|---|---|
| $T_0$ | 296 K |
| $\beta$ | $3.3 \times 10^{-4}$ eV · m$^{0.5}$ |
| $A$ | $1.7 \times 10^{-9}$ A · K$^{-2}$ |
| $D$ | 5 μm |
| $R_{\text{th}}$ | $2.5 \times 10^{8}$ K · W$^{-1}$ |
| $C_{\text{p}}$ | 5.5 nF |
| $C_{\text{E1}}$ | 15 pF |
| $R_{\text{E1}}$ | 27 $k\Omega$ |
| $C_{\text{E2}}$ | 0.5 fF |
| $R_{\text{E2}}$ | 1 $k\Omega$ |
| $R_s$ | 300 $k\Omega$ |
| $\phi$ | 0.58 eV |
| $R_{osc}$ | 50 Ω |
| $C_{osc}$ | 100 pF |
| $k_B$ | $8.62 \times 10^{-5}$ eV · K$^{-1}$ |
| $I_{ext}$ | 6 $\mu$A |

This model produces abrupt resistive switching during a DC voltage sweep, with a change in device current by several orders of magnitude across the transition, displayed in Figure S21a. This is caused by an electrothermal runaway process due to feedback, and although not physically an IMT it produces similar electrical behavior. Similar NDR behavior can also be achieved with other nonlinear thermally-activated transport such as Poole-Frenkel [12], leading to oscillations. When a DC current sweep is performed (Figure S21b), the positive temperature-current feedback is restricted, and the device produces negative differential resistance (NDR) behavior with accompanying oscillations. When a 600 kΩ heater is included (Figure S21c-d), the switching voltage is reduced and the insulating state current is higher (as the heater provides a parallel conductive path for current to flow), reducing the on/off ratio (the metallic state is limited by the series resistance). Both devices show NDR during a current sweep and oscillate. Hysteresis was not included, for ease of calculations. To study periodic oscillations, the circuit was biased with a constant current source, $I_{ext}$ = 6 μA, which operates the device within its region of NDR.

As the tuning parameters $C_{th}$ and $G_{heater}$ (1/$R_{heater}$) are varied the system can abruptly change between slow ~kHz oscillations and faster ~MHz oscillations, as shown in Figure S22 (a cross section of this plot along fixed $R_{heater}$ produced main text Figure 4c). When there is no heater present, or a very resistive heater that does not produce significant heating (zero or small $G_{heater}$), a reduction in $C_{th}$ (representing the thermal switching volume, which is reduced as the device is scaled) cannot produce faster oscillations. However, if both $C_{th}$ and $R_{heater}$ are sufficiently small, then fast oscillations occur, demonstrating that adding a heater in addition to scaling the device is one way to speed up the system.

We note that an abrupt transition in frequency (a bifurcation) in this model appears to require the presence of two capacitors in the circuit and does not occur when only a single capacitor is included in parallel with the device. The addition of a second capacitor (separated by a resistor from the other capacitor) adds another voltage variable $v_{cap}$, increasing the circuit from a second order (depending on the state variables $T$ and $v_m$) to a third order system. This increased complexity makes the system more susceptible to bifurcations, and the actual devices and experimental set-up certainly had various capacitances. Bifurcations are not specific to this circuit only, and were found to occur in other configurations of the circuit with different values or locations for the contact and parasitic resistances and capacitances. Self-sustained oscillations do not occur when a very small resistor $R_{heater}$ is used, which prevents effective access of the NDR in the $VO_2$ by electrically shorting it. Bifurcations or fast oscillations did not occur when a parallel resistor was electrically included in the circuit but which did not contribute to heating (no term involving $R_{heater}$ in Eqn. S6).

## 6. Video

Video S1 (provided as a separate file) shows the simulated steady state evolution of the electric field and temperature in a nanogap device as the applied voltage is increased (see also Figure 2c,f in the main text). The device dimensions correspond to that of the real device in Figure S1a, which was also used for KPM and SThM measurements. The temperature shown is on the surface of the 35 nm PMMA capping layer, to best compare with SThM, but the $VO_2$ surface temperature is slightly hotter. IMT1 occurs between 10 - 11 V (the $VO_2$ in the gap is insulating for applied voltages $V_S \leq 10$ V and metallic for $V_S \gtrsim 11$ V), and IMT2 occurs beyond 16 V in the simulation.

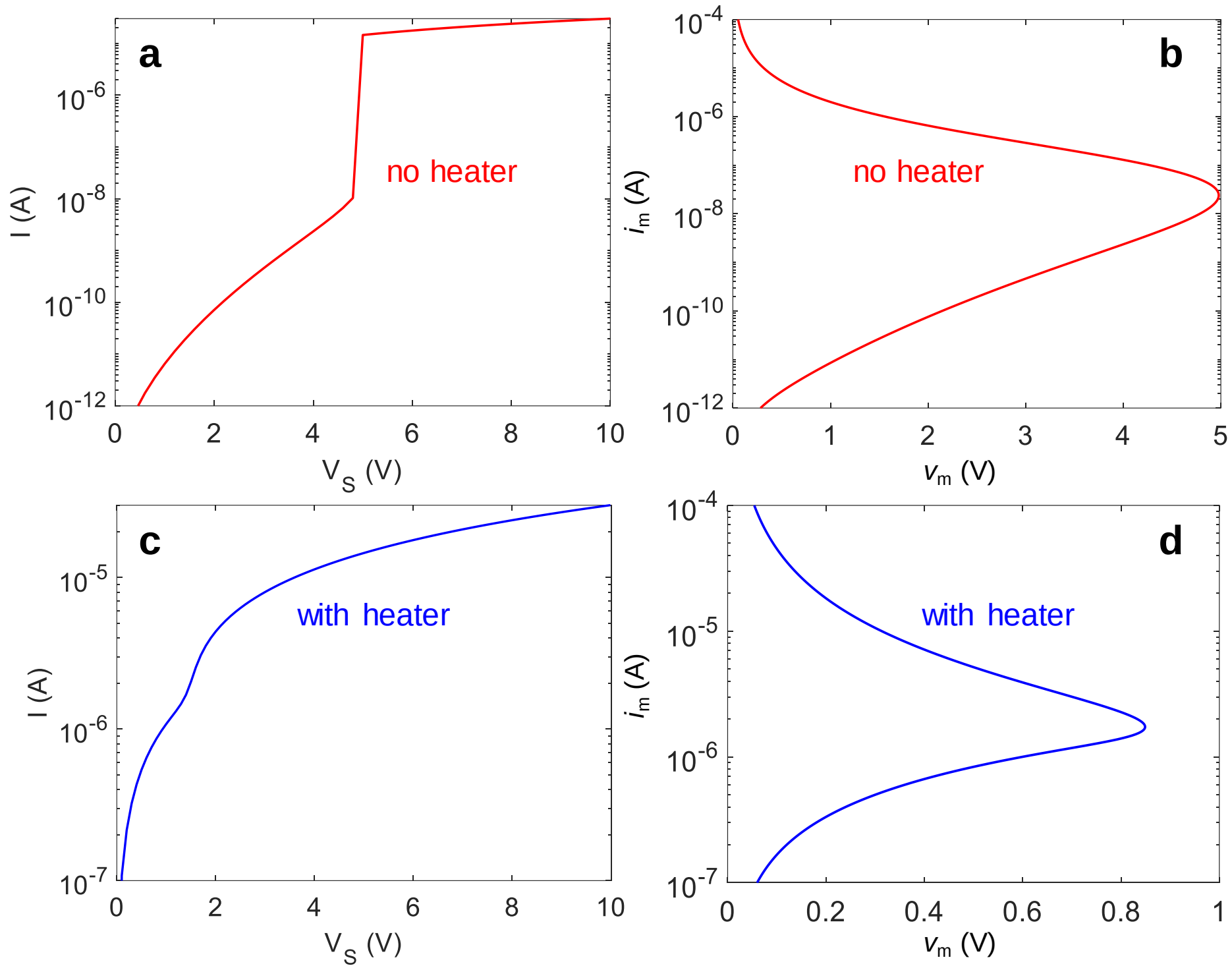

**Figure S22.** DC *I-V* characteristics simulated in SPICE: (a) with a voltage source and no heater present, (b) with a current source and no heater, (c) with a voltage source and $R_{heater}$ = 600 kΩ, and (d) with a current source and $R_{heater}$ = 600 kΩ. Voltage sweeps are plotted as viewed from the source node ($V_S$), and current sweeps as seen at the device ($v_m$).

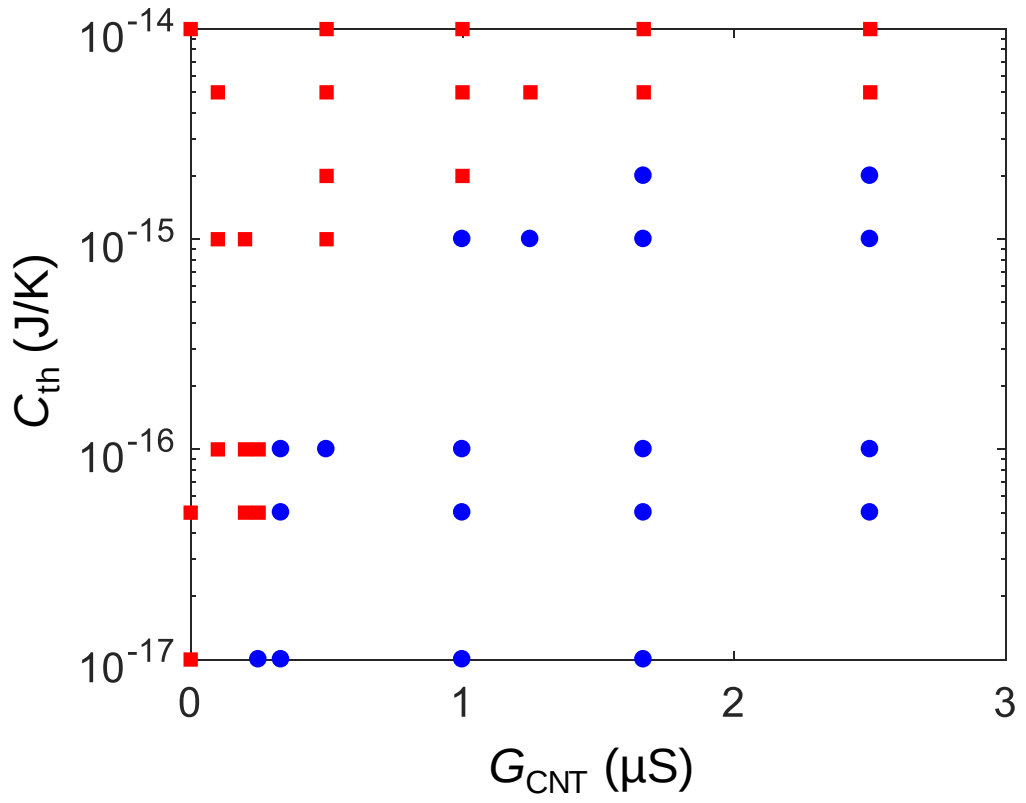

**Figure S23.** Oscillation frequency abruptly transitions between “slow” (red squares, ~kHz) and “fast” (blue circles, ~MHz) depending on the choice of $C_{th}$ and heater conductance $G_{heater}$ ($1/R_{heater}$).

## 7. LT SPICE files

LT SPICE files which include the simulated circuit and compact device model, as well as all parameters, are included as additional supplement files. These simulations were run in version 4.13m of LT SPICE.